\documentclass[%
 amsmath,amssymb,
 prfluids,
]{revtex4-2}

\usepackage{graphicx}
\usepackage{dcolumn}
\usepackage{bm}

\usepackage[utf8]{inputenc}
\usepackage[T1]{fontenc}
\usepackage{mathptmx}
\usepackage{etoolbox}
\usepackage{color}

\def\bea{\begin{equation}}
\def\eea{\end{equation}}

\newcommand{\pdiffn}[3]{\frac{\partial^{#3} #1}{\partial #2^{#3}}}
\newcommand{\pdiff}[2]{\frac{\partial #1}{\partial #2}}

\begin{document}

\preprint{APS/123-QED}

\title[Putting the micro into the macro]{Putting the micro into the macro:\\A molecularly-augmented hydrodynamic model of dynamic wetting applied to flow instabilities during forced dewetting}
\author{J. S. Keeler}
\email{jack.keeler@warwick.ac.uk}
 \affiliation{Mathematics Institute, University of Warwick, Coventry, CV4 7AL, UK}
\author{T. B. Blake}%
 \email{terrydblake@btinternet.com}
\noaffiliation{} 

\author{D. A. Lockerby}
 \email{duncan.lockerby@warwick.ac.uk}
\affiliation{School of Engineering, University of Warwick, Coventry, CV4 7AL, UK}%

\author{J. E. Sprittles}
\email{J.E.Sprittles@warwick.ac.uk}
 \affiliation{Mathematics Institute, University of Warwick, Coventry, CV4 7AL, UK}

\date{\today}

\begin{abstract}
We report a molecularly-augmented continuum-based computational model of dynamic wetting and apply it to the displacement of an externally-driven liquid plug between two partially-wetted parallel plates. The results closely follow those obtained in a recent molecular-dynamics (MD) study of the same problem \citep{toledano2021closer}, which we use as a benchmark. We are able to interpret the maximum speed of dewetting $U^*_{\mathrm{crit}}$ as a fold bifurcation in the steady phase diagram and show that its dependence on the true contact angle $\theta_{\mathrm{cl}}$ is quantitatively similar to that found using MD. A key feature of the model is that the contact angle is dependent on the speed of the contact line, with $\theta_{\mathrm{cl}}$ emerging as part of the solution. The model enables us to study the formation of a thin film at dewetting speeds $U^*>U^*_{\mathrm{crit}}$ across a range of length scales, including those that are computationally prohibitive to MD simulations. We show that the thickness of the film scales linearly with the channel width and is only weakly dependent on the capillary number. This work provides a link between matched asymptotic techniques (valid for larger geometries) and MD simulations (valid for smaller geometries). In addition, we find that the apparent angle, the experimentally visible contact angle at the fold bifurcation, is not zero. This is in contrast to the prediction of conventional treatments based on the lubrication model of flow near the contact line, but consistent with experiment. 
\end{abstract}
\maketitle
\section{\label{sec:level1}Introduction}
Dynamic wetting, the process by which a liquid wets a solid surface, is an important phenomenon that underpins a wide range of both industrial and natural processes, including microfluidics \citep{stone2004}, liquid coating and printing operations \citep{weinstein2004}, petroleum recovery \citep{gerritsen2005model}, plant protection \citep{papierowska2018contact}, ground water hydrology \citep{beatty2010fractional} and biological processes \citep{barthlott2016superhydrophobic}. As such, it presents a multiscale problem. Whilst its origin is at the microscopic scale of the moving contact line, it influences outcomes at very much larger scales. 
However, despite this importance, and consequent research over many decades, there remain fundamental questions about the physics involved and, in particular, the role of solid-liquid interactions at the moving contact line \citep{andreotti2020statics,afkhami2020challenges,semenov2011droplets}.

In wetting studies, solid-liquid interactions are usually quantified in terms of the angle of contact between the liquid and the solid, and its proper description has attracted much attention  \citep{de1985wetting,Blake2006,shikhmurzaev2007capillary,andreotti2020statics}. From hydrostatic and hydrodynamic perspectives, this boundary condition is crucial, as it dictates the shape of the liquid volume.  The way it changes in response to movement of the contact line across the solid surface is, therefore, fundamental to our ability to predict wetting outcomes. Nevertheless, the description of the true contact angle at a moving contact line remains hotly debated \citep{andreotti2020statics,afkhami2020challenges,semenov2011droplets}. 

\begin{figure}
  \centering
  \includegraphics[scale=0.7,trim=0 0 0 0,clip]{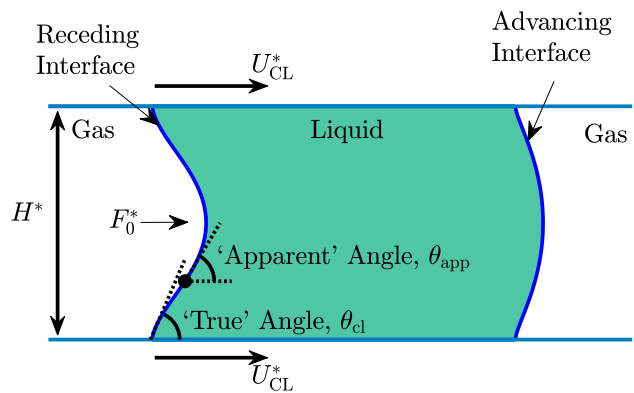}
  \caption{Schematic of a liquid plug between two plates subject to an external forcing $F^*_0$. The angle the receding contact line makes with the plate is the true angle, denoted $\theta_{\mathrm{cl}}$. See figure~\ref{fig:apparent_angle} for a more detailed schematic of the angle measurements.}
  \label{fig:liquid_bridge}
\end{figure}

In continuum models the true contact angle, measured by the tangent of the interface at the solid (see figure~\ref{fig:liquid_bridge}), has to be specified in order to solve the governing equations and is usually considered to be constant and equal to the equilibrium value. The observed dynamics of the apparent contact angle (i.e. the one seen experimentally \citep{wilson2006nonlocal}) is attributed to the `viscous bending' of the interface: this is the so-called `Hydrodynamic' or Cox-Voinov formulation \citep{cox1985part1,Voinov1976}. However, according to the molecular-kinetic theory of wetting (MKT) \citep{Blake1969, Blake1993} and the interface formation model \citep{shikhmurzaev2007capillary}, the true contact angle varies and is dependent on the velocity of the contact line. 

Here, we will show that viscous bending alone is insufficient to capture the effects seen in molecular simulations, where the velocity dependence of the actual contact angle is observed. Therefore, we develop a new combined approach based on the Navier-Stokes continuum paradigm combined with the MKT, (whose formulation is far simpler than the interface formation model, despite the latter's attractive features), and focus it on the canonical dynamic wetting problem of a liquid plug propagating through a channel. In particular, in order to allow for unambiguous comparisons to the results of molecular dynamics on a comparable system, we identify a critical speed at which a flow bifurcation occurs and a thin film is formed. 


\begin{figure}
  \centering
  \includegraphics[scale=0.38]{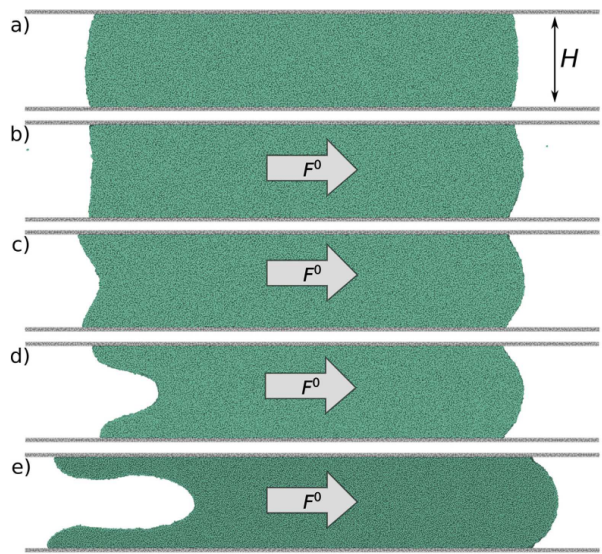}
  \caption{Figure reprinted from \cite{toledano2021closer}, with permission from Elsevier. The panels a) to e) show the liquid plug as a the force $F^*_0$ (in the article asterisks were not used to denote dimensional quantities) becomes successively larger and eventually exceeds the critical value (panels d) and e)) where a thin-film begins to develop. In this MD simulation, the external phase is a vacuum.}
  \label{fig:blake_figure}
\end{figure}

The study of dynamic wetting using molecular simulations has a long history, see review articles \cite{de2008wetting} and \cite{koplik1995continuum}; but here we focus on a recent paper, \cite{toledano2021closer}, that examines both wetting transitions and the behaviour of the contact angle. In this study, large-scale molecular dynamics (MD) is utilised to explore the steady displacement of a water-like liquid plug between two molecularly-smooth solid plates under the influence of an external driving force $F^*_0$ (see figure~\ref{fig:liquid_bridge} for the geometry). The study used a coarse-grained model of water and an atomistic Lennard-Jones model for the solid plates. The general behaviour observed as $F^*_0$ was increased, and hence the liquid-plug's speed was raised, is depicted in figure~\ref{fig:blake_figure}. Notably, it was reported that both the `true', dynamic contact angle at the contact line, $\theta_{\mathrm{cl}}$, and a larger-scale `apparent' angle, $\theta_{\mathrm{app}}$, are dependent on the contact-line velocity $U_{\mathrm{cl}}^*$ for the receding and advancing interfaces. Henceforth, unless otherwise stated, when we refer to a `contact angle' we mean the `true' contact angle. We also note that quantities labelled with an asterisk correspond to dimensional physical quantities and those without to dimensionless quantities.

In the MD study, the apparent angle was measured at the system scale by a method that mimics typical measurements of it in macroscopic experiments, where the precise details of the true contact angle's dynamics remain hidden, as they occur on such small length scales \citep{Dussan1979,hoffman1975study,Blake2006}.  By varying the solid-liquid affinity (i.e. the solid's wettability), it was possible to investigate the influence of the equilibrium contact angle $\theta_0$  on the results.  For all $\theta_0$, $\theta_{\mathrm{cl}}$  was found to be velocity dependent in a manner consistent with the MKT of dynamic wetting \citep{Blake1969, Blake1993}. However, $\theta_{\mathrm{app}}$  diverged from $\theta_{\mathrm{cl}}$ as $F^*_0$  was increased, especially at the receding contact line, in a way that closely followed the Voinov equation \citep{Voinov1976}: 
\bea
\theta_{\mathrm{app}}^3 = \theta_{\mathrm{cl}}^3 + 9\,Ca\log(L^*/L_m^*),
\label{cox_voinov}
\eea
where $Ca=\mu^* U_{\mathrm{cl}}^*/\gamma^*$ is the capillary number based on the contact-line speed $U_{\mathrm{cl}}^*$, dynamic viscosity $\mu^*$ and surface tension $\gamma^*$, and $L^*$ and $L_m^*$ are suitably-chosen macroscopic and microscopic length scales.  For each $\theta_0$, there was a critical receding contact-line velocity $U_{\mathrm{crit}}^*$ and contact angle $\theta_{\mathrm{crit}}$ at which $\theta_{\mathrm{app}}$ {became small} and the receding meniscus deposited a liquid film on the plates. This value could then be used in \eqref{cox_voinov}, assuming $\theta_{\mathrm{app}}\approx 0$, to fix $L^*/L^*_m$ and hence reliably predict $\theta_{\mathrm{cl}}$ at both the advancing and receding contact lines. This result is significant, as $\theta_{\mathrm{cl}}$  is not usually experimentally accessible and the fact that it varies with $U_{\mathrm{cl}}^*$ poses questions for hydrodynamic interpretations of dynamic wetting.  The result also shows that the critical condition for film deposition encodes crucial information about the hydrodynamics.  

The existence of a critical wetting speed has been investigated thoroughly using hydrodynamic models in a range of geometries, including those associated with coating flows \citep{kumar2015liquid}, and plate withdrawal \citep{snoeijer2008thick}, among others. In \cite{keeler2021stability}, both receding and advancing contact line problems were investigated for a coating flow and the stability of the solutions near the critical speed was quantified using a dynamical systems method. Here, our focus will be on the receding contact line, as this is where the first bifurcation will occur. Previous studies have shown that as $Ca$ increases, the receding contact line will attain a steady state provided $Ca<Ca_{\mathrm{crit}}$, where $Ca_{\mathrm{crit}}$ is a critical capillary number that is a function, at the very least, of $\theta_{\mathrm{cl}}$, the slip length and the viscosity ratio of the liquid and gas phases \citep{snoeijer2007part1,snoeijer2006avoid,eggers2004forced,cox1985part1,keeler2021stability}, but, if $Ca>Ca_{\mathrm{crit}}$, a thin-film develops with thickness dependent on $Ca$ \citep{snoeijer2006avoid,keeler2021stability}. Using a lubrication model, $Ca_{\mathrm{crit}}$ can be approximated when the slip-length is small relative to the film height \citep{eggers2005existence}, by considering a small-$Ca$ asymptotic analysis and using the key assumption that $\theta_{\mathrm{app}} = 0$ at the critical point. However, in a nano-geometry, as considered here, we will see that this assumption is not valid and the resulting small-$Ca$ asymptotic analysis does not extend to this regime.

In this paper we will develop a hydrodynamic model based on the Navier-Stokes paradigm to calculate steady states and transient behaviour of the liquid plug scenario considered in ~\cite{toledano2021closer}. An essential aspect of this model is that the true angle, $\theta_{\mathrm{cl}}$, has to be specified at the junction of the liquid, gas and solid phases. In many previous studies where a Navier-Stokes model is used (see, for example ~\cite{kamal2019dynamic,sprittles2012finite,liu2019predictions,vandre2012delay,liu2016assist,liu2016surfactant,liu2017mechanism,vandre2013mechanism}) $\theta_{\mathrm{cl}}$ is assumed to be constant; but motivated by the results of ~\cite{toledano2021closer} we relax this assumption and adopt a model that determines $\theta_{\mathrm{cl}}$ as a function of $Ca$ and the static contact angle $\theta_0$ based on the MKT. Notably, the model remains hydrodynamic throughout, in contrast, for example, to ~\cite{hadjiconstantinou1999hybrid}, and the molecular-augmentation comes entirely through the contact-angle formula. This approach was also considered for macroscopic flows in ~\cite{dodds2012dynamics}, where the static contact angle, $\theta_0$ and slip length are independent parameters. However, motivated again by \cite{Blake2015} and \cite{toledano2020hidden}, we will make use of a correlation between slip length and $\theta_0$ that reduces the number of parameters that are required. This correlation is based on an assumption, borne out by MD simulations, that the mechanism of slip between a liquid and a solid is the same across all parts of the solid-liquid interface, including the contact line. 


The article is ordered as follows. In \S~\ref{sec:hydro_model} we describe the system of equations used to model the liquid plug based on the Navier-Stokes (NS) equations. In addition to the NS equations, in \S~\ref{sec:qp} we discuss asymptotic results, based on a Quasi-Parallel (QP) lubrication approach adapted from \cite{eggers2005existence}, that will be relevant here. By calculating numerical solutions of the governing equations using a finite-element framework, we will then show in \S~\ref{sec:system} that the critical speed of wetting for the entire liquid plug is dependent on the receding contact line and not influenced by the advancing contact line. We will also discuss the method for calculating the apparent angle. Next, in \S~\ref{sec:steady} we will show how augmenting the NS equations with an MKT variable-angle constraint predicts the existence of a critical $Ca$, and that as the wettability is varied the values of $Ca_{\mathrm{crit}}$ match favourably with the MD data in \cite{toledano2021closer}, in contrast to the predictions of the fixed-angle model. In addition, we demonstrate how $\theta_{\mathrm{cl}}$ and $\theta_{\mathrm{app}}$ vary with the slip length and thus provide an estimate of $L^*/L_m^*$ for the liquid plug system, which shows excellent agreement with the MD simulations. Further, in \S~\ref{sec:thinfilm} we examine time-dependent behaviour when $Ca>Ca_{\mathrm{crit}}$ so that a thin film develops, whose height obeys a Landau-Levich law. Finally, in \S~\ref{sec:large_scale}, having validated the system in the liquid nano plug geometry, we exploit our computational framework to explore larger-scale systems, which are beyond the scope of MD simulations. By examining systems where the physical size is orders of magnitude larger than the nano-channel studied in \cite{toledano2021closer}, we will show that the {dimensionless thickness of the film remains constant, for a fixed $Ca$.} 

\section{\label{sec:hydro_model} Molecular-Augmented Hydrodynamic Model\protect\\}

We will now describe the hydrodynamic model. We shall discuss the full system, based on the Navier-Stokes equations and then describe the two different system formulations; the pressure-driven problem and the force-driven problem, as well as the numerical method and the different computational domains. 

\subsection{Fully Nonlinear System}

\begin{figure*}
  \centering
  \includegraphics[scale=0.40,trim=270 0 270 0,clip]{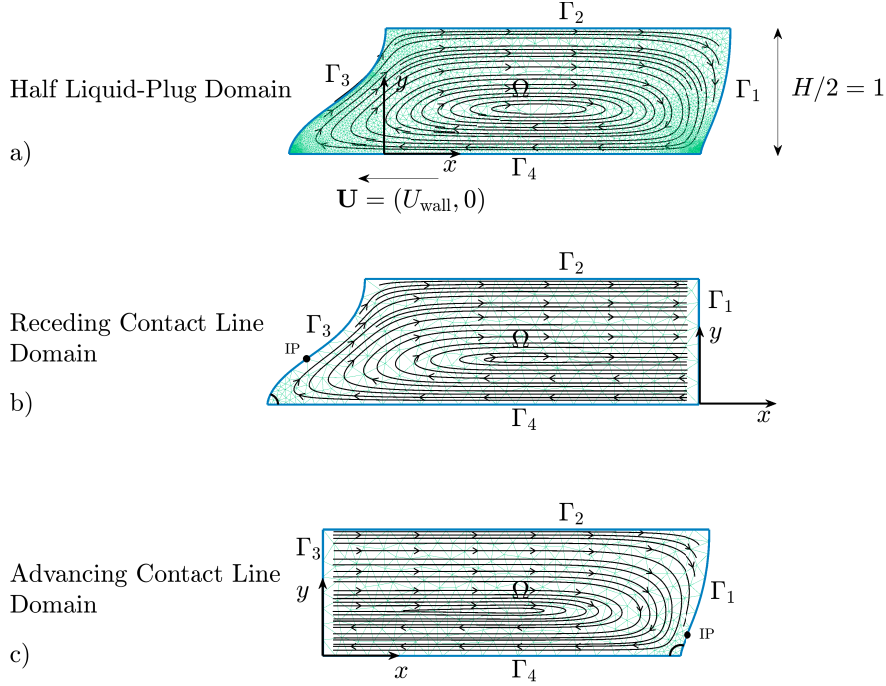}
	\caption{The computational domain with streamlines and computational elements in the background. (a) Half Liquid Plug Domain - the upper boundary, $\Gamma_2$ is a symmetry boundary and $\Gamma_4$ is a moving wall, so we are computing the system in a frame of reference that moves with the liquid. b) Receding contact line (RCL) domain where, instead of a free-surface at $\Gamma_1$, we impose parallel flow which significantly reduces the computational burden. c) Advancing contact line (ACL) domain domain. In b) and c) the circular markers on the free-surfaces indicate the location of the inflection point, denoted IP. Parameter values are $Ca = Ca_{\mathrm{crit}} =  0.31$, $\lambda = 0.1$, $\theta_{\mathrm{cl}} = \pi/2$.}
  \label{fig:mesh_domain}
\end{figure*}

To mimic the molecular simulations, we model the liquid-bridge system as a two-dimensional flow between two parallel plates as illustrated in figure~\ref{fig:liquid_bridge} and detailed in figure~\ref{fig:mesh_domain}(a). A finite liquid region fills the channel bounded by two rigid plates that are separated by a distance $H^*$. {We solve in a frame of reference that moves with the plug, with the
walls moving with velocity $U_{\mathrm{wall}}^*$. The exact formulation depends on the domain and problem that we consider (i.e. pressure-driven or body-force driven), details of which we discuss later.} We nondimensionalise all lengths using the half-height, $H^*/2$, all velocities using $U_{\mathrm{wall}}^*$, all pressures by $\mu^* U_{\mathrm{wall}}^*/(H^*/2)$ all timescales by $(H^*/2)/U_{\mathrm{wall}}^*$ { and the body force by $\mu^* U_{\mathrm{wall}}^*/(H^*/2)^2$.} As in other studies \citep{vandre2013mechanism,vandre2012delay,liu2019predictions,liu2016assist,liu2016surfactant,liu2017mechanism,sprittles2013finite,sprittles2011viscous1,sprittles2011viscous2}, we apply the Stokes-flow approximation, \eqref{eq1}-\eqref{eq2}, so that the Reynolds number, $\rho^*U_{\mathrm{wall}}^*H^*/\mu^*$, is assumed to be negligibly small;  simple estimates confirm that this is appropriate for the nano-system. We neglect the influence of gravity and assume that that the gas phase can be modelled as a vacuum (as seen in figure~\ref{fig:blake_figure}, there are no molecules in the gas phase). A typical computational domain is shown in figure~\ref{fig:mesh_domain}(a). On the moving wall ($y=0$)  we apply a Navier-slip condition, \eqref{eq3}, and, therefore, introduce a dimensionless slip-length, $\lambda$. The MD simulations in figure~\ref{fig:blake_figure} indicate the flow is symmetric around the centreline of the channel and hence we introduce a symmetry wall at $y=1$, labelled $\Gamma_2$, where we set the vertical component of velocity to be zero, apply zero tangential stress, and let the horizontal velocity be determined as part of the solution. As well as the fluid velocity field, $\textbf{u}(t,\textbf{x})$, and pressure, $p(t,\textbf{x})$, which depend on the dimensionless time, $t$, and the position, $\textbf{x}$ of the interfaces, denoted $\textbf{R}_{\mathrm{adv}}=(x_{a}(t,s),y_{a}(t,s))$ and $\textbf{R}_{\mathrm{rec}}=(x_{r}(t,s),y_{r}(t,s))$ respectively, are also unknowns in the problem and functions of $t$ and the arclength, $s$, as measured from the contact point. These are found using dynamic and kinematic conditions on both free surfaces. The governing equations and boundary conditions then become
\begin{align}
  &0=\:-\nabla p + \nabla^2\textbf{u} + \textbf{F}, \qquad\textbf{x}\in\Omega,&\mbox{Conservation of Momentum},
  \label{eq1}\\
  &\nabla\cdot\textbf{u} =\: 0,\qquad\textbf{x}\in\Omega,&\mbox{Incompressibility},\label{eq2}\\
  &\lambda ( \boldsymbol{\tau} \cdot \textbf{n})\cdot\textbf{t}, =\: (\textbf{u} - \textbf{U})\cdot \textbf{t},\qquad\textbf{x}\in\Gamma_{4},&\mbox{Navier-Slip Condition},\label{eq3}\\
&\textbf{u}\cdot\textbf{n} =\:0, \qquad\textbf{x}\in\Gamma_{4},&\mbox{No-Penetration Conditon},\label{eq3a}\\
	&\textbf{u}\cdot\textbf{n} =\:0,\,(\boldsymbol{\tau}\cdot\textbf{n})\cdot \textbf{t} =0, \qquad\textbf{x}\in\Gamma_{2},&\mbox{Symmetry Condition},\label{eq4}\\
	&\boldsymbol{\tau}\cdot\textbf{n} = \frac{1}{Ca}\kappa \textbf{n},\qquad\textbf{x}\in\Gamma_{1}\cup\Gamma_{3},&\mbox{Dynamic Condition (Rec. and Adv.)},\label{eq5}\\
&\pdiff{\textbf{R}_{\mathrm{adv}}}{t}\cdot\textbf{n} = \textbf{u}\cdot\textbf{n},\qquad\textbf{x}\in\Gamma_{1},&\mbox{Kinematic Condition (Adv.)},\label{eq6}\\
&\pdiff{\textbf{R}_{\mathrm{rec}}}{t}\cdot\textbf{n} = \textbf{u}\cdot\textbf{n},\qquad\textbf{x}\in\Gamma_{3},&\mbox{Kinematic Condition (Rec.)}, \label{eq7}
\end{align}
where $\textbf{n}$, and $\textbf{t}$ are the vectors normal and tangential, respectively, to the appropriate boundaries denoted $\Gamma_i$, and $\kappa$ is the curvature of the corresponding interface. The plate speed $\textbf{U} = (U_{\mathrm{wall}},0)^T$ and $\lambda = \lambda^*/(H^*/2)$ is the dimensionless slip length. The body force is $\textbf{F} = (F,0)^T$, where $F$ is a set constant. The stress tensor $\boldsymbol{\tau}$ is defined as
\bea
\boldsymbol{\tau} = -p\textbf{I} + \left(\nabla \textbf{u} + (\nabla\textbf{u})^T\right),
\eea
where $\textbf{I}$ is the identity matrix. We shall refer to the system described by \eqref{eq1}-\eqref{eq7} as the `Half Liquid-Plug' problem. 

\subsection{Contact-Angle Models}
The system is not well-posed unless a contact angle is specified between the free-surfaces and the horizontal plates. For the symmetry boundary we set $\theta(s = L) = \pi/2$, but the dynamic contact angle, $\theta_{\mathrm{cl}}$, can be freely chosen and depends on the wettability of the solid. 

\begin{figure}
  \centering
  \includegraphics[scale=0.6,trim=0 0 0 0,clip]{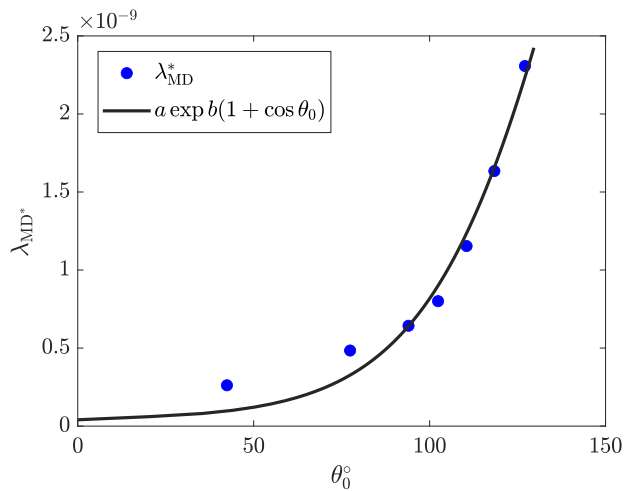}
  \caption{The slip length dependence on the static angle. The markers are the MD data obtained from \cite{toledano2021closer} and the solid line is the curve fit given from \eqref{lambda_theta_0} with $b = -2.342$ and $a = 5.656\times 10^{-9}$.}
  \label{fig:slip_curve_fit}\end{figure}

The simplest approach is to specify a constant equilibrium contact angle, i.e.
\bea
\theta_{\mathrm{cl}} = \mbox{Const}.\qquad \mbox{Constant Angle Equation}
\label{constant_theta}
\eea
However, as is well known, the molecular-kinetic theory predicts $\theta_{\mathrm{cl}}$ to be dependent on the speed of the contact line. As shown in Appendix~\ref{app:MD_theory}, for the receding contact lines of interest here, this dependence may be written in the linearised form  
\bea
\overline{Ca} =\frac{\lambda}{\delta}\left(\cos(\theta_0) - \cos(\theta_{\mathrm{cl}})\right),\qquad  \mbox{Variable Angle Equation}
\label{variable_theta}
\eea
where $\delta$ is a dimensionless parameter that corresponds to the width of the three-phase zone, i.e. the contact line viewed at the molecular scale, and $\overline{Ca}$ is the relative velocity of the contact line to the wall speed, i.e.
\bea
\overline{Ca} = Ca\left(\textbf{U}\cdot \textbf{e}_x - \left.\pdiff{x}{t}\right|_{s = 0}\right),
\label{rel_ca}
\eea
where $\textbf{e}_x$ is a unit vector in the $x$ direction. Equation \eqref{variable_theta} is the linearised form of the theory, which will be valid for the system considered in this article. {Furthermore, it has long been recognised that a relationship must exist between the slip length and the equilibrium contact angle, e.g., \cite{tolstoi1952,barrat1999influence,priezjev2007rate}.  Here, motivated by the results of \cite{toledano2021closer} and the theory described in Appendix~\ref{app:MD_theory},}we consider the relationship
\bea
\lambda_{\mathrm{MD}}^* = a\exp\left[b(1 + \cos(\theta_0)\right],\quad 0<\theta_0<\pi,
 \label{lambda_theta_0}
 \eea
 where $a$ and $b$ are fitting parameters and $\lambda_{\mathrm{MD}}^*$ is the physical slip-length derived from the MD data in \cite{toledano2021closer}. We will assume that $\lambda_{\mathrm{MD}}^*$ is independent of the physical channel height, $H^*$, and therefore in our non-dimensionalisation $\lambda = 2\lambda_{\mathrm{MD}}^*/H^*$. To investigate the nano-channel used in \cite{toledano2021closer}, where $H^* = 20.2 \mbox{nm}$, the different values of $\theta_0$ will yield dimensionless slip-lengths in the range $\lambda \sim 0.02$ to $0.2$. Alternatively, as we will show in \S~\ref{sec:large_scale}, by varying $\lambda$, and keeping $\theta_0$ fixed, we can investigate the effects of varying the physical channel height $H^*$ to larger systems. We emphasise that there is a one-to-one correspondence between $\lambda_{\mathrm{MD}}^*$ and $\theta_0$, and hence we are free to prescribe either quantity and use \eqref{lambda_theta_0} to determine the other. In physical experiments it is more practical to find $\theta_0$, which can readily be measured, and then determine $\lambda_{\mathrm{MD}}^*$, which is more difficult to measure experimentally. Figure~\ref{fig:slip_curve_fit} shows the fit of this function to the MD data from \cite{toledano2021closer}. We call the system of equations described in \eqref{eq1}-\eqref{eq7} augmented with the constant angle formula, \eqref{constant_theta}, the Constant-Angle (CA) model, while when augmented with the variable angle formula, \eqref{variable_theta} and \eqref{lambda_theta_0}, we call it the Variable-Angle (VA) model. 
 

\subsection{Pressure-driven and Force-driven Problems}

We now discuss the two different types of problems, i.e. the pressure-driven and force-driven problems. {The MD simulations in \cite{toledano2021closer} is a force-driven problem, but pressure-driven problems are relevant in many practical situations, for example, coating flows, \citep{liu2019predictions}.}

In Pouseille flow, without a free-surface, it is easy to show that a pressure-driven problem can be equivalent to a force-driven one. With a free-surface however, this is not true and each case has to be considered separately. For steady calculations, in both types of problems, the set of equations are ill-posed unless we specify the volume of the liquid-plug. To remove this issue, for pressure-driven flow, we impose a normal stress on $\Gamma_1$:
\bea
\boldsymbol{\tau}\cdot\textbf{n} = p_{\mathrm{out}}\textbf{n},\qquad\textbf{x}\in\Gamma_1,
\label{eq8}
\eea
and let the value of $p_{\mathrm{out}}$ be determined implicitly by a condition on the overall volume of the liquid-plug, which corresponds to the computational area of the domain. We set $F=0$ and solve in a frame of reference that moves such that the walls are non-stationary in the translating frame ($U_{\mathrm{wall}} = -1$). 

In contrast, for the steady force-driven problem, we set $U_{\mathrm{wall}} = -1$ and $p_{\mathrm{out}} = 0$, but now let $Ca$ be determined implicitly by a volume constraint. {In both cases, to overcome the translational invariance, we also have to pin a point on the boundary which depends on the domain of the problem (as discussed below).}

For time-dependent problems, whether pressure-driven or force-driven, the volume constraint is unnecessary, as equation~\eqref{eq2} ensures that volume is conserved. Instead, we impose a position constraint for the reduced domains (see below) that determines $p_{\mathrm{out}}$ or $Ca$, depending on the problem.

\subsection{Numerical Method}

The complete system of equations are discretised and solved using a finite-element method and the open-source \texttt{oomph-lib} package \citep{heil2006oomph}, as described in \cite{keeler2021stability}. An unstructured triangular mesh is used which is treated as a pseudo-elastic body, so that changes to the unknown free-surface can be facilitated and the mesh can adapt to capture regions of high velocity or pressure gradients, for example near the contact point. We use a ZZ error estimator, which measures the continuity of the rate of strain in each element, to identify elements that require refinement or unrefinement \citep{Zienkiewicz1992}. As a typical example for the time-dependent calculations, with $Ca =0.5$ and $\lambda = 0.01$, elemental areas range from $\sim10^{-3}-10^{-1}$ to accommodate a maximum ZZ error of $10^{-3}$. 


\subsection{`Half' and `Quarter' Domains}

{We can simplify the complexity of the half liquid-plug domain further by solving in two separate `quarter' {domains}, each having only one free surface; thus significantly reducing the number of triangular elements required, see figures~\ref{fig:mesh_domain}(b) and (c). To facilitate this, we replace the free-surface (and corresponding dynamic and kinematic boundary conditions) at one end of the computational domain ($\Gamma_1$ for the advancing contact line and $\Gamma_3$ for the receding contact line) with an imposed normal stress (i.e. equation \eqref{eq8}) and parallel flow condition 
\bea
v = 0,\qquad\textbf{x}\in\Gamma_1 (\mbox{Receding})\quad \mbox{or} \quad \textbf{x}\in \Gamma_3 (\mbox{Advancing}).
\label{eq9}
\eea
In these quarter {domains}, the imposed pressure, $p_{\mathrm{out}}$, is determined implicitly by ensuring the volume per unit length (i.e. the area) of the liquid domain is constant (as in the half liquid-plug {domain}), so that in each quarter {domain} we are solving in a frame of reference with a fixed volume. We note that in the quarter domains we impose \eqref{eq8} and \eqref{eq9} in both steady and time-dependent calculations. These quarter domain simulations will be referred to as the `Receding Contact Line' and `Advancing Contact Line' {domains} (see figures~\ref{fig:mesh_domain}(b) and (c)), respectively, abbreviated to RCL and ACL in the rest of the paper. {The origin is different in each of these domains and corresponds to the pinned position of the steady and time-dependent problems.} To illustrate the benefit of this reduction, the number of elements required in the computation of figure~\ref{fig:mesh_comparison} for the half-liquid plug is $\sim 4000 - 8000$, but the number for the RCL is $\sim 400$. {The reason for the $\sim$ 90\% reduction in elements is because the the pressure gradients are not as severe near the receding contact line, compared to the advancing one.} In the next section, we shall show that the dynamics of the whole system and the prediction of a critical $Ca$ are dominated by the receding contact line. Thus, computation of the full half liquid-plug problem, where both the advancing and receding interface are calculated, is not necessary in order to to find the first flow bifurcation and is computationally inefficient when compared to the reduced RCL domain. }
\newline

We emphasise that the main aim of this study is to investigate the VA model, and not to make a thorough investigation of the differences between pressure-driven and force-driven flow, as the VA model can be applied independently of the problem. Thus, in the results that follow, we mainly consider pressure-driven flow, except when we make a direct comparison with the MD simulations. For the latter, we present force-driven results, as clearly specified.  In addition, as we shall show in \S~\ref{sec:finding_ca_crit}, the choice of domain is also independent of the problem (i.e. pressure-driven or force-driven), so we choose the domain that is most important to the flow-bifurcation and which is also the simplest, computationally.

\section{Reduced Governing Equations (Quasi-Parallel System)}\label{sec:qp}
We shall now discuss a reduced evolution PDE model, the so-called `Quasi-Parallel' system, before finally obtaining asymptotic results that will help predict the value of $Ca_{\mathrm{crit}}$.

For the receding contact line, the flow near the contact line is approximately parallel, c.f. figure~\ref{fig:thin_film_quiver}, and we can exploit this to reduce the Navier-Stokes equations to a simpler system which requires unknowns only on the fluid interface. As well as the parallel-flow assumption, we assume the horizontal coordinate is approximately the arclength, i.e. $x\approx s$, so the full expression for the curvature can be used, not the linearized form as used in conventional lubrication models (see, for example \cite{eggers2005existence}) and long-wave models (see, for example \cite{snoeijer2006free}). 

Following \cite{jacqmin2004onset}, \cite{sbragaglia2008wetting} and \cite{vandrethesis}, we let $\theta$ be the angle the interface makes to the horizontal (see figure~\ref{fig:liquid_bridge}), $h=y_r(s)$ be the height of the interface and $s$ be the arclength coordinate measured from the contact line. Using conservation of mass and the kinematic condition on the free-surface, the governing equation for the fluid pressure gradient, $\pdiff{p}{s}$, may be written as \citep{snoeijer2006avoid}
\bea
  \pdiff{h}{t} + \pdiff{Q}{s} = 0,\qquad Q = \pdiff{}{s}\left(-\frac{1}{3}\pdiff{p}{s}h^2(h + 3\lambda) - (\textbf{U}\cdot\textbf{e}_x)h\right),
\label{lub_eqn}
  \eea
  where $\textbf{U}\cdot\textbf{e}_x = -1$ for the RCL. The unknown pressure gradient, $\pdiff{p}{s}$, can then be expressed in terms of the exact curvature by differentiating the normal stress balance w.r.t $s$, i.e.
\bea
\frac{1}{Ca}\pdiffn{\theta}{s}{2} = \pdiff{p}{s}.
\label{curv_eqn}
\eea
In order to solve \eqref{curv_eqn}, we require two conditions on $\theta$. At the contact line, $s=0$, we implement equation \eqref{variable_theta}:
\bea
\overline{Ca} = \frac{\lambda}{\delta}\left(\cos(\theta_0) - \cos(\theta_{\mathrm{cl}})\right),
\eea
and at the symmetry wall, $s=L$, we set $\theta = \pi/2$, where $L$ is the overall length of the interface. The shape of the interface can then be recovered by solving
\bea
\pdiff{x}{s} = \cos(\theta),\qquad \pdiff{h}{s} = \sin(\theta).
\label{pos_eqn}
\eea
Each of these equations requires a single condition, so we set $x(s=0) = y(s=0) = 0$ (choosing the contact line to be at the origin). Finally we note that the length of the interface, $L$, and hence the size of the domain, $s$, are not known \textit{a priori}. To determine $L$ we scale the independent variable $s$, so that $\xi = Ls$ and $\xi = [0,1]$. The total length of the interface, $L$, can then be determined by the additional constraint that
\bea
y(\xi = 1) = 1.
\eea
To solve this system of equations, we choose to discretise the spatial derivatives using finite-differences and then the system of equations are solved numerically using Newton's method. We remark that we exclusively concentrate on the steady results of the {pressure-driven} QP system and do not solve the time-dependent problem, as this is better suited to the full nonlinear system.


\subsection{Asymptotics}
We now briefly describe and adapt the analysis of \cite{snoeijer2012theory} and \cite{eggers2005existence} to find an asymptotic expression for $Ca_{\mathrm{crit}}$. We will not repeat their analysis except for the parts where it differs from the situation we examine here. In both of these previous works gravitational effects are included and the liquid domain is unconfined, whereas we neglect gravity and the system is confined. They also considered only steady solutions, so that time-derivatives in the problem can be ignored.  

The matched asymptotics methodology of \cite{snoeijer2012theory} and \cite{eggers2005existence} is to determine an inner solution, say $h = h_{\mathrm{inner}}$, for small $Ca$, that is valid close to the contact line, i.e. when $s/\lambda \sim \mathcal{O}(1)$, and an outer solution, $h = h_{\mathrm{outer}}$ say, that is valid far away from the contact line, i.e. when $s\sim \mathcal{O}(1)$. To determine an unknown constant in the outer solution, the inner and outer solutions have to match in a crossover region. This matching procedure yields an equation, for an arbitrary unknown, $\theta_{\mathrm{app}}$, which is the angle the outer interface makes with the horizontal. For the ACL domain, $\theta_{\mathrm{app}}$ is finite for all values of $Ca$, and thus, in these asymptotic limits at least, there is no critical point for the ACL domain. However, in the RCL domain $\theta_{\mathrm{app}}$ can only be calculated up to a critical value of $Ca$, this value being interpreted as $Ca_{\mathrm{crit}}$. 

In our problem, for a confined geometry and in the absence of gravity, the inner region analysis near the contact line is identical to the case considered in \cite{snoeijer2012theory} and \cite{eggers2005existence}. {In \cite{eggers2005existence} the effects of gravity are present at first-order in the outer solution.} The outer solution, for small $Ca$, is found by expanding the unknowns as a power series in $Ca$. We can write a leading-order outer solution of \eqref{lub_eqn} to \eqref{pos_eqn} as
\begin{equation}
	\begin{split}
	&\theta_{\mathrm{outer}} = \theta_{\mathrm{app}} + \kappa_s s, \qquad
h_{\mathrm{outer}} = 1-\frac{1}{\kappa_s}\cos\left(\theta_{\mathrm{app}} +  \kappa_s s\right),\\
	&x_{\mathrm{outer}}  = \frac{1}{\kappa_s}\left[\sin\left(\theta_{\mathrm{app}} +  \kappa_s s\right) - 1\right] + X_{\mathrm{rec}},
	\end{split}
	\label{outersol}
	\end{equation}
where $\kappa_s = (\pi - 2\theta_{\mathrm{app}})/2L$ is the curvature, $X_{\mathrm{rec}}$ is the meniscus rise (c.f. figure~\ref{fig:mesh_comparison}) and $\theta_{\mathrm{app}}$ is an undetermined constant. The outer solution in \eqref{outersol} describes a sector of a circle with centre $(X_{\mathrm{rec}} - r,1)$ and $r = 1/\kappa_s$ that makes an angle $\theta_{\mathrm{app}}$ to the horizontal at $s=0$. Examining the geometry (see dashed curve in figure~\ref{fig:apparent_angle}(a)) gives
$
\kappa_s = \sin(\pi/2 - \theta_{\mathrm{app}}),
$
so as $\theta_{\mathrm{app}}\to 0$, $\kappa_s\to 1$ and hence the interface is a semi-circle of radius 1. In particular we have that
\bea
L\to \pi/2,\qquad X_{\mathrm{rec}}\to 1,\qquad\mbox{as}\qquad\theta_{\mathrm{app}}\to 0,
\label{thetaapp0limit}
\eea
When the outer solution, described in \eqref{outersol}, is matched to the inner solution, as described in \cite{eggers2005existence}, we obtain an expression for $\theta_{\mathrm{app}}$ 
\bea
\frac{\theta_{\mathrm{app}}}{\theta_{\mathrm{cl}}^3} = -\frac{2^{2/3}3^{1/3}Ca^{1/3}\mathrm{Ai}'(z_1)}{\mathrm{Ai}(z_1)},
\label{thetaapp}
\eea
where $\mathrm{Ai}(z)$ is an Airy function of the first kind. We note this is the same for an unconfined geometry with gravity as considered in \cite{eggers2005existence}. In addition, $Ca_{\mathrm{crit}}$ satisfies the same expression as in \cite{eggers2005existence} but has a factor of $2^{1/3}$ in the denominator of the logarithm term, i.e.
\begin{equation}
        Ca_{\mathrm{crit}} = \frac{\theta_{\mathrm{cl}}^3}{9}\left[\log\left(\frac{Ca_{\mathrm{crit}}^{1/3}\theta_{\mathrm{cl}}}{3^{2/3}\cdot2^{1/3}\mathrm{Ai}^2(z_{\mathrm{max}})\lambda\pi}\right)\right]^{-1},\qquad z_{\mathrm{max}} = -1.0188..
\label{ca_crit}
\end{equation}
These results are valid for a constant contact angle model but we can easily extend them to our variable angle formula by expanding \eqref{variable_theta} in powers of $Ca$, for $Ca\ll 1$, and then $\theta_{\mathrm{cl}}$ in the above expression is just the static angle $\theta_{0}$. However, we will show that the full expression for $\theta_{\mathrm{cl}}$ in \eqref{variable_theta} will be required in the formula \eqref{ca_crit} for it to compare favourably with the numerical results. In the sections that follow, the expressions in \eqref{thetaapp0limit},\eqref{thetaapp} and \eqref{ca_crit} will be compared with the numerical solutions.

\section{System Measurements and Parameters}\label{sec:system}
In this section we describe the methods used to determine $Ca_{\mathrm{crit}}$ from the numerical calculations of the fully nonlinear and quasi-parallel systems. In addition we also discuss, in detail, the methodology of identifying the value of $\theta_{\mathrm{app}}$, and compare two different methods.

\subsection{Finding the Critical $Ca$} \label{sec:finding_ca_crit}
\begin{figure}
  \centering
  \includegraphics[scale=0.31,trim=0 0 0 0,clip]{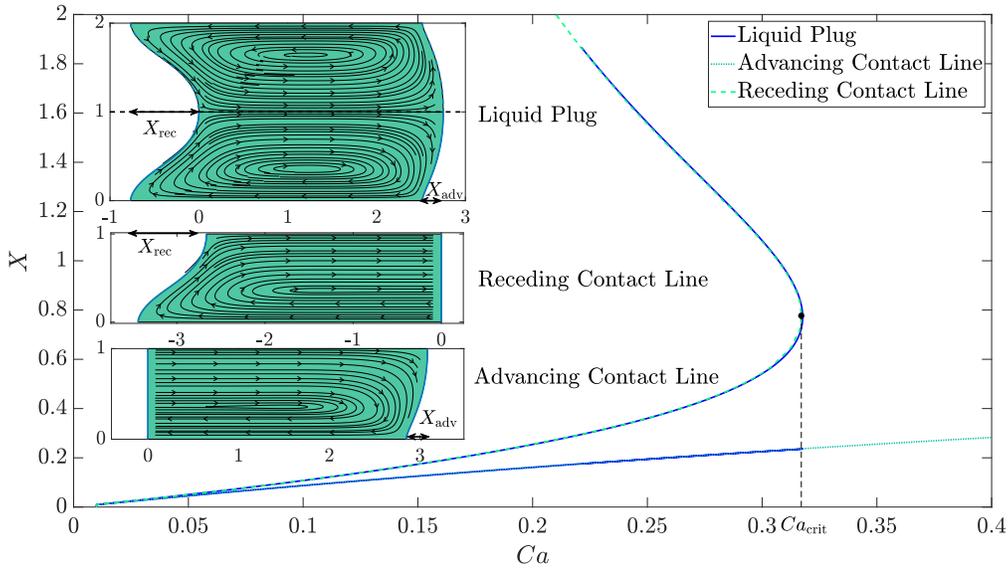}
	\caption{The steady solution space mapped in the $(Ca,X)$ plane when $\theta_{\mathrm{cl}} = \mbox{Const} = \pi/2$. $X_{\mathrm{adv}}$ and $X_{\mathrm{rec}}$ are the horizontal distances of the interface at the two plates for the advancing and receding case respectively. The solid curves represent the half liquid-bridge problem, while the broken lines indicate the `quarter' problems where the advancing and RCL are calculated separately. The limit point for the RCL indicates the threshold beyond which no steady states exist, denoted by $Ca_{\mathrm{crit}}$, and corresponds to the critical $F^*_0$ in the MD simulations of \cite{toledano2021closer}. The inset diagrams correspond to the parameter values $Ca = Ca_{\mathrm{crit}} \approx  0.31$, $\lambda = 0.2$}
  \label{fig:mesh_comparison}
\end{figure}

We now describe how we determine the critical $Ca$ computationally. We note that this methodology is valid for both the fully nonlinear and quasi-parallel systems. Initially, we shall assume a constant $\theta_{\mathrm{cl}}$, i.e. we impose \eqref{constant_theta}, and for simplicity we shall assume that $\theta_{\mathrm{cl}} = \pi/2$. In the pressure-driven problem, by varying $Ca$ and subsequently solving the steady set of equations, we can trace the state of the system by recording the horizontal distance between where the interface meets the moving wall and the symmetry wall, which we denote $X_{\mathrm{rec}}$ and $X_{\mathrm{adv}}$ for the receding and advancing contact lines, respectively {(the same can be achieved in the force-driven problem by increasing the value of $F$ and finding the maximum value of $Ca_{\mathrm{crit}}$)}. Figure~\ref{fig:mesh_comparison} shows the resulting solution curves of $X_{\mathrm{rec}}$ (upper curve) and $X_{\mathrm{adv}}$ (lower curve) plotted against $Ca$. The solid lines indicate solutions of the half liquid-plug problem, while the broken lines are solutions of the corresponding quarter RCL and ACL domains. 

There are a number of important features of these solution curves. The RCL solution curve experiences a limit point (or fold bifurcation) where the curve turns around and the corresponding steady solution becomes unstable. The consequence of this is that the value of $Ca$ where this critical point occurs marks the limiting threshold for which stable (i.e. those which can be experimentally realised) steady solutions exist, and it is, therefore, natural to associate this value with $Ca_{\mathrm{crit}}$. It is important to emphasise that the limit point occurs only for the RCL in this setup (i.e. a liquid-vacuum system). Furthermore, we note that the curves of the half liquid-plug problem completely overlap the curves for the RCL and ACL domains, the only distinction being that the quarter ACL solution curves are able to continue past $Ca_{\mathrm{crit}}$, as no unstable ACL is observed. The critical point of the full nonlinear system coincides with the fold bifurcation of the RCL, and so in order to understand the dynamics of the system before and after criticality, we only need to consider the RCL; thus, significantly reducing the computational demands.

\subsection{Measuring the Apparent Angle}
\begin{figure}
  \centering
  \includegraphics[scale=0.28,trim=100 0 0 0,clip]{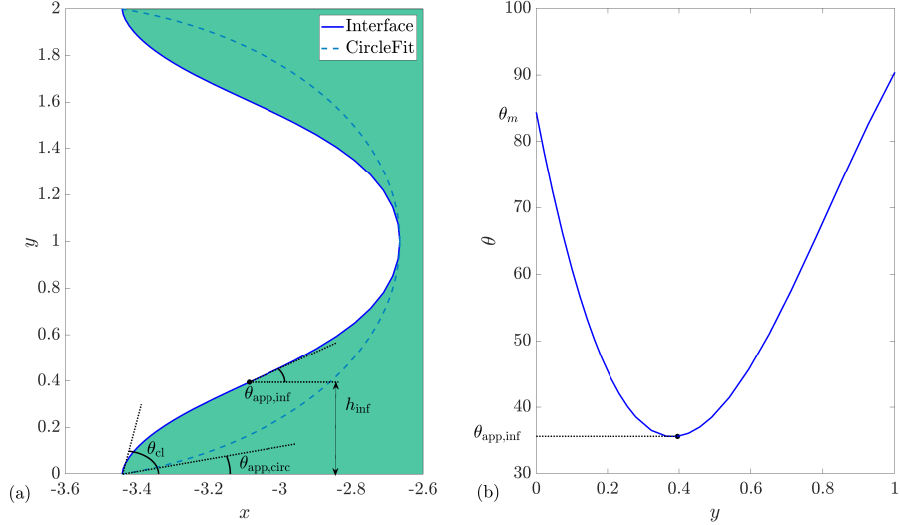}
  \caption{The alternative definitions of the apparent angle. In (a) the $\theta_{\mathrm{app,circ}}$ is defined as the angle a fitted circle makes with the bottom plate. $\theta_{\mathrm{app,inf}}$ is defined as the minimum angle the interface makes with the horizontal, as measured anti-clockwise, and corresponds to the inflection point of the interface, where the curvature, $\kappa = 0$. In (b) we plot the interface angle as a function of $y$ which demonstrates that $\theta_{\mathrm{app,inf}}$ can be calculated as the minimum value of $\theta$.}
  \label{fig:apparent_angle}
\end{figure}

The precision with which the dynamic contact angle can be measured experimentally is limited by the resolution of the method used \citep{Dussan1979}.  This is usually of the order of a few micrometres, and for optical measurements can be no better than the diffraction limit.  Thus, accurate measurement of the true contact angle is not possible, though some progress has been made \citep{Chen2014}.  A common approach is to fit a curve to the image of the interface and measure its tangent at its point of intersection with the solid surface.  Alternatively, the interface can be assumed to have a quasi-equilibrium shape (e.g. a spherical cap) from which the angle may be deduced via an appropriate formula \citep{hoffman1975study,Dussan1979,Chen1995,Lhermerout2019}.  Neither approach has the ability to resolve significant changes in curvature very close to the contact line, such as those observed in the MD simulations (see figure~\ref{fig:blake_figure}), which occur whenever the true contact angle differs significantly from the apparent angle. Therefore, to a greater or lesser extent, the measured angle inevitably depends on the method used to measure it, and simply represents the slope of the interface at some arbitrary distance from the contact line. This is the reason why these angles are commonly described as `apparent’.  

In the MD study, \cite{toledano2021closer}, two methods were investigated to evaluate $\theta_{\mathrm{app}}$ in a systematic way that was consistent with experiment, despite the very small scale of the system.  For both, multiple snapshots were averaged to account for thermal noise.  Since the menisci of the liquid-plug are cylindrical at rest, the methods were based on circular fits to the liquid surface.  The first approach was to estimate the slope of the interface at the point of its inflection, as shown in figure~\ref{fig:apparent_angle}.  This was achieved by fitting the arc of a circle to the central 50\% of the meniscus (i.e. well away from the inflections) and measuring the slope of the arc at its points of intersection with planes parallel to the solid surfaces and passing through the inflections.  The method appealed as being consistent with the asymptotic matching procedure used in hydrodynamic treatments of dynamic wetting \citep{Voinov1976,cox1985part1}.  

The second approach was to mimic experiment more directly by measuring the tangents to a circular arc defined by upper and lower contact lines and passing through the apex of the meniscus at its mid-point, as shown in figure~\ref{fig:apparent_angle}.  This procedure is commonly used to measure the dynamic contact angle in capillary systems \citep{Dussan1979}; and because the positions of the three defining points could be measured more accurately from the simulations than the locations of the inflections, this was the method adopted.  It gave advancing angles a few degrees smaller than the those found at the inflection points, but the receding angles were indistinguishable within simulation limits.  The method was also used in a recent numerical study of microscopic and apparent contact angles \citep{omori2017apparent}. 

Similarly, in the present paper we calculate $\theta_{\mathrm{app}}$ in two different ways consistent with those adopted in \cite{toledano2021closer}. In the left panel of figure~\ref{fig:apparent_angle} the liquid-gas interface is shown by a solid line and the arc of a circle that is tangent to the interface at the line of symmetry by a dashed line. We define $\theta_{\mathrm{app,circ}}$ as the angle the circle makes with the horizontal as shown in figure~\ref{fig:apparent_angle}. This definition, used in \cite{toledano2021closer}, is useful if the position of the liquid-gas interface is not well defined.

Alternatively, and again as considered in \cite{toledano2021closer}, we can define $\theta_{\mathrm{app,inf}}$ as the angle the interface makes with the horizontal at the inflection point of the curve, see Figure~\ref{fig:apparent_angle}. In this definition we measure the angle along the curve using the identity
\bea
\theta \equiv \mbox{atan}\left(\frac{y'(s)}{x'(s)}\right)
\label{theta_eq}
\eea
and then find the minimum value that $\theta$ takes as a function of $x$, see right panel of Figure~\ref{fig:apparent_angle}. This corresponds to where the curvature is zero and the interface has an inflection point. The value of $y$ at the inflection point is denoted $h_{\mathrm{inf}}$, which will be commented on later in the paper. This approach of measuring $\theta_{\mathrm{app}}$ is more amenable to FEM calculations, because \eqref{theta_eq} can be calculated easily as the position of the interface is well-defined and was the approach used in \cite{liu2019predictions}, \cite{vandre2012delay}, \cite{liu2016assist,liu2016surfactant,liu2017mechanism} and \cite{vandre2013mechanism}.

\section{Steady Results\label{sec:steady}}

In this section we will describe the steady solution space of the RCL using and comparing the CA model, where $\theta_{\mathrm{cl}}$ is held constant, and the VA model, where $\theta_{\mathrm{cl}}$ is determined by \eqref{variable_theta}. First, using parameter values that are representative of the values in \cite{toledano2021closer}, we shall compute steady solutions in the {pressure-driven} VA model and present a bifurcation diagram that demonstrates that the fold bifurcation, which represents the critical speed of dewetting, still exists when using the VA model and predicts the value obtained in \cite{toledano2021closer}.  Then we shall vary $\theta_0$ to investigate the effect of wettability on the value of $Ca_{\mathrm{crit}}$ and compare this directly with the results of \cite{toledano2021closer} {using the pressure-driven and force-driven problem}. Finally, we compare the predictions of the continuum model with the results previously obtained by applying the Cox-Voinov law to the data derived from the MD simulations in \cite{toledano2021closer}
\subsection{Bifurcation Diagram and General Features}

\begin{figure}
  \centering
  \includegraphics[scale=0.3,trim=0 0 0 0,clip]{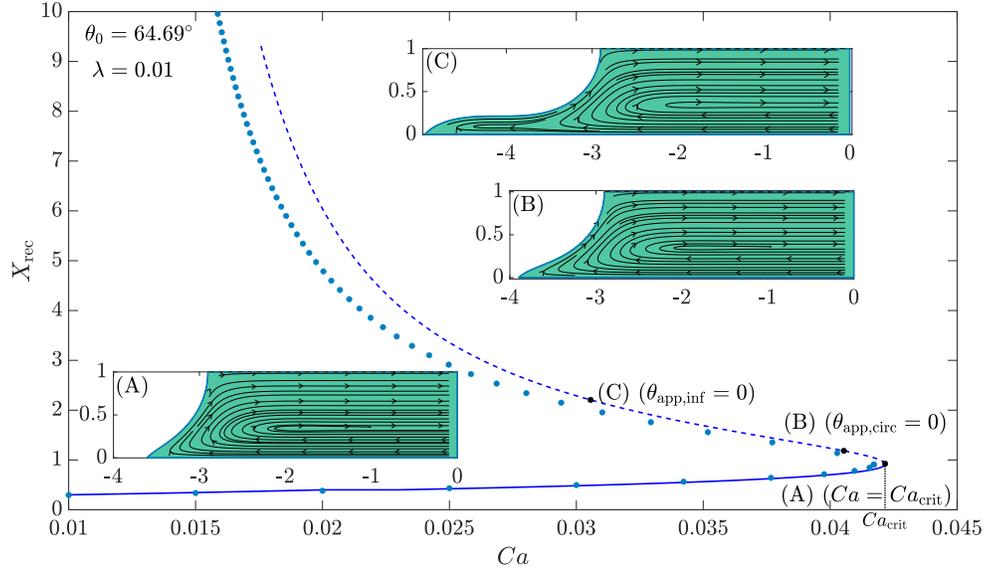}
  \caption{The steady solution structure for $\theta_0 = 64.7^{\circ}$ and $\lambda = 0.02$. The solid/dashed curves indicate the stable/unstable branches of the full VA system and the solid circular markers indicate the solution using the QP approach. The inset profiles show the steady solution interface and domain when (A) $Ca = Ca_{\mathrm{crit}}$, (B) $\theta_{\mathrm{app,circ}} = 0$ and (C) $\theta_{\mathrm{app,inf}} = 0$.} {Their locations are indicated by solid black markers on the main curve.}
  \label{fig:bifurcation_diagram_lambda_0_01}
\end{figure}

We now focus our attention to the VA model and discuss steady solutions and the critical point. We emphasise that in this model we require only the static angle, $\theta_0$, the width of the three-phase zone, $\delta$, and the capillary number, $Ca$, as specified parameters so that a steady solution can be computed.

Figure~\ref{fig:bifurcation_diagram_lambda_0_01} shows the steady solution space of the RCL domain by plotting $X$ against $Ca$, as calculated numerically. The solid and dashed curves indicate, respectively, the stable and unstable solution branches of the full system and the circular markers indicate the solution branch of the QP system.  The inset diagrams show streamline patterns and the interface position at (A) the critical point, (B) when $\theta_{\mathrm{app,circ}} = 0$ and (C) when $\theta_{\mathrm{app,inf}} = 0$. There are a number of interesting features that are worth commenting on. First, we note that even though $\theta_{\mathrm{cl}}$ is now dependent on $Ca$, the limit point still occurs. We also remark on the close agreement of the QP system and the full nonlinear system; the QP system does remarkably well in approximating the limit point for this particular value of $\theta_0$, although the curves diverge as $X_{\mathrm{rec}}$ increases. 

The solution where $\theta_{\mathrm{app,circ}}=0$ is significantly closer on the bifurcation curve to the limit point than where $\theta_{\mathrm{app,inf}} = 0$. In fact, as seen from the inset interface profiles, the interface when $\theta_{\mathrm{app,inf}} = 0$ (label (C)) is already significantly deformed and approaching a thin-film, whereas the profile when $\theta_{\mathrm{app,circ}} = 0$ (label (B)) more closely matches the interface at the limit point (label (A)). This result is interesting, as in many works, e.g. \cite{eggers2005existence}, $Ca_{\mathrm{crit}}$ is defined as occuring when $\theta_{\mathrm{app}} = 0$. This is strictly valid in the regime $\lambda\to 0$ and we note that in our geometry the slip length and $Ca$ take moderate values, and therefore we find that $\theta_{\mathrm{app}}\neq 0$ at $Ca_{\mathrm{crit}}$.

The consequence of our findings is that it is not unreasonable to use the definition of $\theta_{\mathrm{app,circ}} = 0$ as a lower bound on the critical capillary number when analysing outcomes in experiments and MD simulations where the existence of a smooth bifurcation curve is hidden, {including smaller geometries whose dimensions are comparable to those of the slip length}. Furthermore, at the limit point, we expect the dynamics of the system to be very slow, as the leading eigenvalue of the linear stability problem will be close to zero \citep{keeler2021stability}. Thus in any experimental/MD setup the time frame may not be large enough to guarantee that a steady state is being approached or whether a thin-film is about to develop. Therefore, we conclude that while the position of the critical point is a well-defined threshold for the critical capillary number in calculations involving a deterministic hydrodynamic model, for experimental and MD results using the definition of $Ca_{\mathrm{crit}}$ as the location where $\theta_{\mathrm{app,circ}} = 0$ may be operationally acceptable.  
\begin{figure}
  \centering
  \includegraphics[scale=0.35,trim=0 0 0 0,clip]{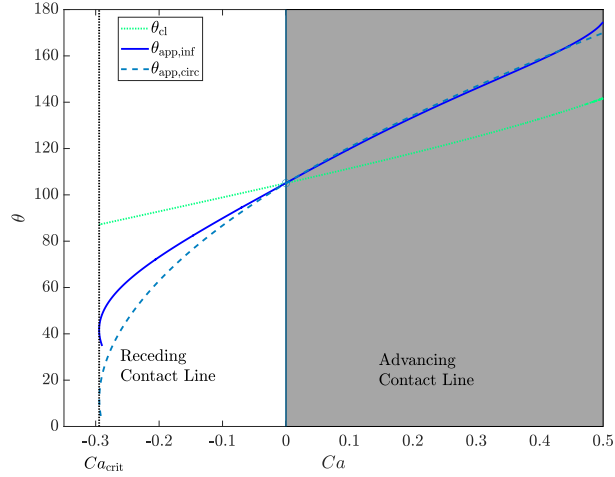}
	\caption{a) The measured angles as a function of $Ca$ when $\lambda = 0.2,\theta_0 = 105.1^\circ$. In this figure we adopt the convention of \cite{toledano2021closer} where for the RCL domain $Ca$ is negative.}
  \label{fig:advancing_receding_bif_diagram}
\end{figure}
The difference between $\theta_{\mathrm{app,circ}}$ and $\theta_{\mathrm{app,inf}}$ for both the RCL and ACL is shown in figure~\ref{fig:advancing_receding_bif_diagram} for values of $\lambda = 0.2$ and $\theta_0 = 105^\circ$. As $Ca_{\mathrm{crit}}$ is approached in the RCL domain the apparent angle tends to zero in a way such that $\theta_{\mathrm{app,circ}}<\theta_{\mathrm{app,inf}}<\theta_{\mathrm{cl}}$. For the ACL domain (the shaded region in the figure), the order of the inequalities is reversed, although there is less of a distinction between $\theta_{\mathrm{app,circ}}$ and $\theta_{\mathrm{app,inf}}$. We should also mention here that in the original MD study \citep{toledano2021closer}, the behaviour of the cosine of the advancing contact angle was significantly non-linear over the range of velocities investigated. As a result, the difference between $\theta_{\mathrm{cl}}$ and $\theta_{\mathrm{app,circ}}$ was significantly smaller than that depicted in figure~\ref{fig:advancing_receding_bif_diagram}, where the linear form of the MKT, \eqref{variable_theta}, is used throughout.

\subsection{Behaviour of $Ca_{\mathrm{crit}}$: VA and CA Model vs. MD.}

\begin{figure}
  \centering
  \includegraphics[scale=0.35,trim=0 0 0 0,clip]{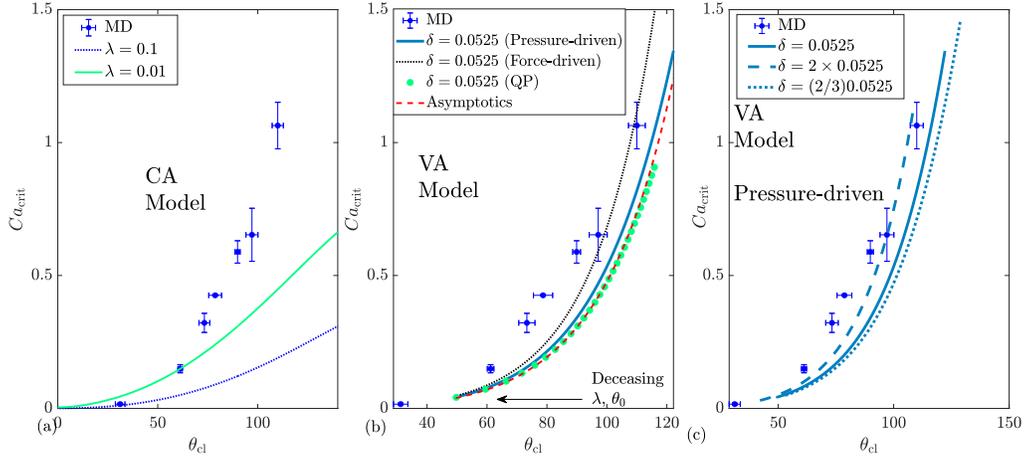}
  \caption{In each chart, the markers with error bars are the MD data obtained from \cite{toledano2021closer}. (a) The critical $Ca$ as a function of the true angle, $\theta_{\mathrm{cl}}$ for different values of $\lambda$ using the constant $\theta_{\mathrm{cl}}$ model given in \eqref{constant_theta}.  (b) The critical $Ca$ as a function of the true angle, $\theta_{\mathrm{cl}}$ for the pressure-driven (solid line) and force-driven (dotted line) problems, QP system (circular markers) and the asymptotics given by \eqref{ca_crit} (dashed line). (c) The critical $Ca$ as a function of the true angle, $\theta_{\mathrm{cl}}$ for different values of $\delta$. The value of $\delta = 0.0525$ corresponds that obtained from the simulations in \cite{toledano2021closer}.}
  \label{fig:loci_critical_ca}
\end{figure}

A critical test of both the CA and VA model is how well it is able to predict $Ca_{\mathrm{crit}}$ when compared with the MD. For the CA model we specify $\lambda$ and $\theta_{\mathrm{cl}}$ and then $Ca_{\mathrm{crit}}$ can be calculated using the method described in \cite{keeler2021stability} to find the fold bifurcation. Figure~\ref{fig:loci_critical_ca}(a) shows the location of $Ca_{\mathrm{crit}}$ as $\theta_{\mathrm{cl}}$ is varied for $\lambda = 0.02$ and $\lambda = 0.2$; these values roughly corresponding to the lower and upper bounds of the slip-length in the MD calculations. As can be seen from the figure, the comparison with the MD data is poor, with $Ca_{\mathrm{crit}}$ underestimated. This provides motivation to implement a VA model where $\theta_{\mathrm{cl}}$ is a function of $Ca$ and $\lambda$.

A much more convincing result is obtained when we apply the VA model. Figure~\ref{fig:loci_critical_ca}(b) shows $Ca_{\mathrm{crit}}$ plotted against $\theta_{\mathrm{cl}}$. The curves represent the loci of the critical point as $\lambda$ and, therefore, $\theta_{0}$, are varied.  Note that $\lambda$ and $\theta_0$, are expressly linked by expression \eqref{lambda_theta_0}.  Here, we have chosen a range of $\lambda$ that matches the MD simulations in \cite{toledano2021closer}; the only parameter we have to specify is $\delta$ in \eqref{variable_theta}. The solid/dotted lines are for the {for the pressure-driven/force-driven problems, respectively,} with $\delta = 0.0525$, the value from the MD simulations; see appendix~\ref{app:MD_theory}. The solid markers are the QP data and the dashed line is the asymptotics described by \eqref{ca_crit}. 

Remarkably, the QP model replicates the full nonlinear system and the simple asymptotic formula shows excellent agreement with the QP model, despite this being in a regime when the slip-length, and indeed $Ca$, are not particularly small. We stress that here the asymptotic formula uses the full $Ca$-dependent formula for $\theta_{\mathrm{cl}}$.


Evidently, the VA theory captures the same qualitative behaviour exhibited by the MD simulations and is much better at predicting $Ca_{\mathrm{crit}}$ than the CA model {and the force-driven problem has a slightly better quantitative fit}. For the value of $\delta$ obtained from \cite{toledano2021closer}, the VA under predicts $Ca_{\mathrm{crit}}$ for the same value of $\theta_0$, but if we make $\delta$ larger the comparison becomes more favourable as shown by the dashed line in figure~\ref{fig:loci_critical_ca}(c). The uncertainty in selecting the appropriate basis for the measurement of $\delta$ from the simulations is discussed in appendix~\ref{app:MD_theory}. Values of $\delta$ larger than that used here are certainly compatible with the data, depending on the criteria used to define the three-phase zone. This uncertainty is compounded by the inevitable thermal noise in the MD results, despite averaging the data over long periods relative to the simulation timescale, which means that there is an inherent difficulty in obtaining the exact steady state at the critical point in the MD simulations. This would mean that the critical point from the MD simulations should be treated as a lower bound, rather than a precise value.

\begin{figure}
  \centering
  \includegraphics[scale=0.35,trim=0 0 0 0,clip]{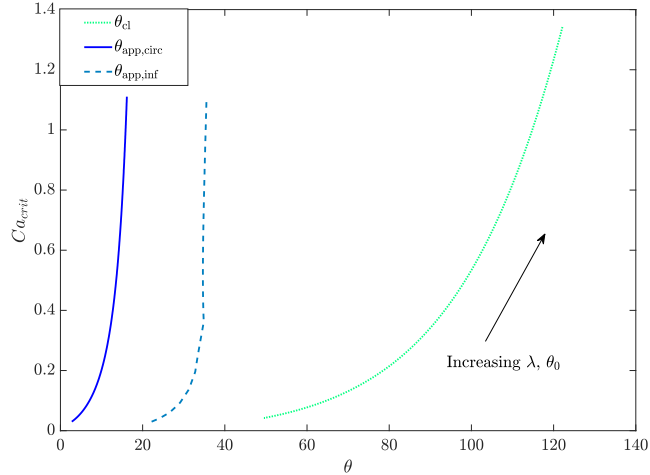}
  \caption{The critical $Ca$ as a function of the true angle for $\delta=0.0525$ with the values of $\theta_{\mathrm{app,circ}}$ and $\theta_{\mathrm{app,Inf}}$ also shown. Note that $\theta_{\mathrm{app,circ}}$ only approaches zero as $\lambda \to 0$}
  \label{fig:loci_critical_ca_with_app_angle}
\end{figure}
As well as measuring $\theta_{\mathrm{cl}}$ we also measure $\theta_{\mathrm{app,inf}}$ and $\theta_{\mathrm{app,circ}}$ at the limit point. Figure~\ref{fig:loci_critical_ca_with_app_angle} shows three curves that represent the $Ca_{\mathrm{crit}}$ as a function of 1) $\theta_{\mathrm{app,circ}}$, 2) $\theta_{\mathrm{app,inf}}$ and 3) $\theta_{\mathrm{cl}}$. We observe that $\theta_{\mathrm{app,circ}}$ at $Ca_{\mathrm{crit}}$ never exceeds $20^{\circ}$ which is consistent with experimental studies of the apparent angles at receding contact lines \citep{Redon1991,deGennes1986,Brochard1992,Rio2005}, where the data shows that $\mbox{d}\theta_{\mathrm{app}}/\mbox{d} Ca$ diverges rapidly as $Ca\to Ca_{\mathrm{crit}}$, and is replicated in the VA model as shown in figure~\ref{fig:advancing_receding_bif_diagram}. In these previous studies apparent angles approaching zero are reported only on surfaces that exhibit little or no contact angle hysteresis \citep{Lhermerout2019}. Hysteresis implies the presence of surface imperfections, such as roughness or heterogeneity.  These cause local fluctuations in contact-line velocity, which may trigger film deposition prematurely.  This might be the reason why \cite{Rio2005} report that angles below about 30 degrees were inaccessible.  The alternative possibility is that dewetting systems may become intrinsically unstable at some value of $\theta_{\mathrm{app}}>0$, as demonstrated here. The fact that $\theta_{\mathrm{app}}$ is itself an artificial construct and dependent on the method of observation adds further uncertainty to the interpretation of experimental data. Nevertheless, we comment that the results from the VA model are consistent with these experimental observations. 

\begin{figure}
  \centering
  \includegraphics[scale=0.45,trim=0 0 0 0,clip]{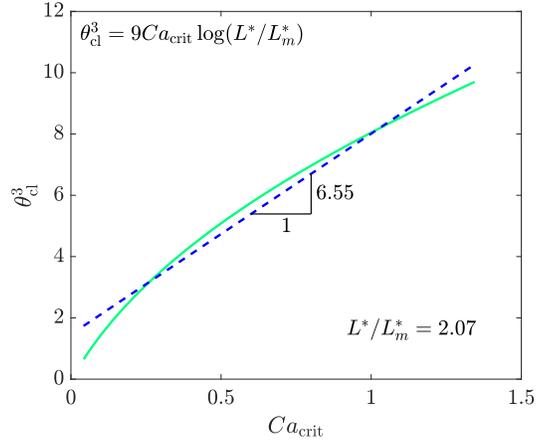}
	\caption{Exploiting the Cox-Voinov law. The solid curve is $\theta_{\mathrm{cl}}^3$ against $Ca_{\mathrm{Crit}}$ as calculated using the full nonlinear model. The dashed line is a line of best fit with gradient corresponding to $L^*/L_m^* = 2.07$.}
  \label{fig:l_lm}
\end{figure}

We shall now use the Cox-Voinov law, equation \eqref{cox_voinov}, to help further rationalise the MD results. If we interpret $\theta_{\mathrm{app}}$ as $\theta_{\mathrm{app,circ}}$, the approach taken by \cite{toledano2021closer}, we can make the approximation that $\theta_{\mathrm{app,circ}} = 0$ at the critical capillary number and then \eqref{cox_voinov} reduces to 
\bea
\theta_{\mathrm{cl}}^3 = 9\,Ca_{\mathrm{crit}}\log\left(L^*/L_m^* \right).
\eea
In figure~\ref{fig:l_lm}, we plot $\theta_{\mathrm{cl}}^3$ as a function of $Ca_{\mathrm{crit}}$, based on the solutions at $Ca_{\mathrm{crit}}$ for each value of $\theta_0$. We can estimate the value of $L^*/L_m^*$ by approximating the curve as a straight line and measuring the slope. It is clear from the figure that a straight line is not wholly appropriate, but its slope gives an approximation of $L^*/L_m^* = 2.07$, which compares favourably with the equivalent approximation from \cite{toledano2021closer} of 2.06 (denoted $L/L_m$ in their study). This is further direct evidence that the numerical results of the model closely replicate the MD simulations.

\section{Time-dependent Results: Thin-Film Formation\label{sec:thinfilm}}

\begin{figure}
  \centering
  \includegraphics[scale=0.3,trim=0 0 0 0,clip]{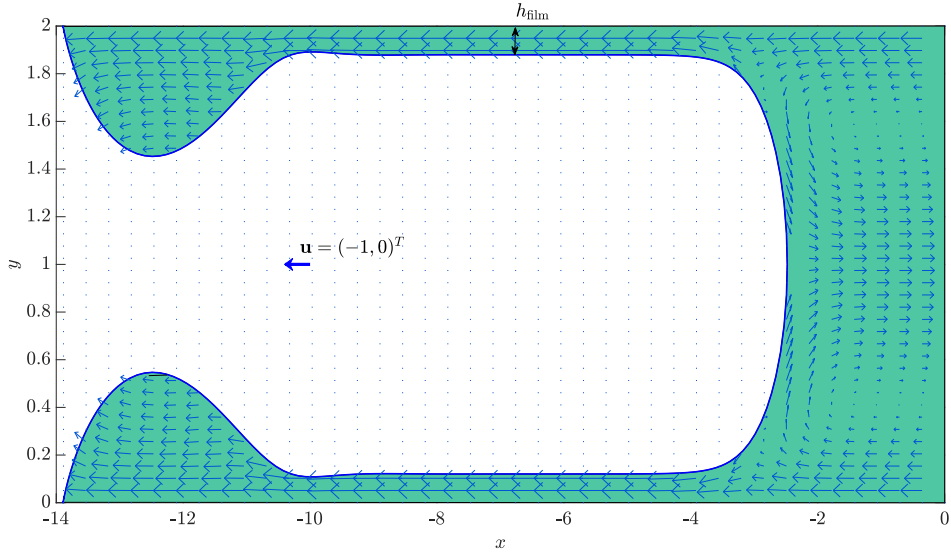}
	\caption{Quiver plot of the thin-film. $Ca = 0.05,\lambda = 0.02,\theta_0 = 64.7^{\circ},t = 19.9$. The arrows indicate the relative size of the local velocity vector field. The blue arrow indicates the scale of a unit vector.}
  \label{fig:thin_film_quiver}
\end{figure}

\begin{figure}
  \centering
  \includegraphics[scale=0.3,trim=0 0 0 0,clip]{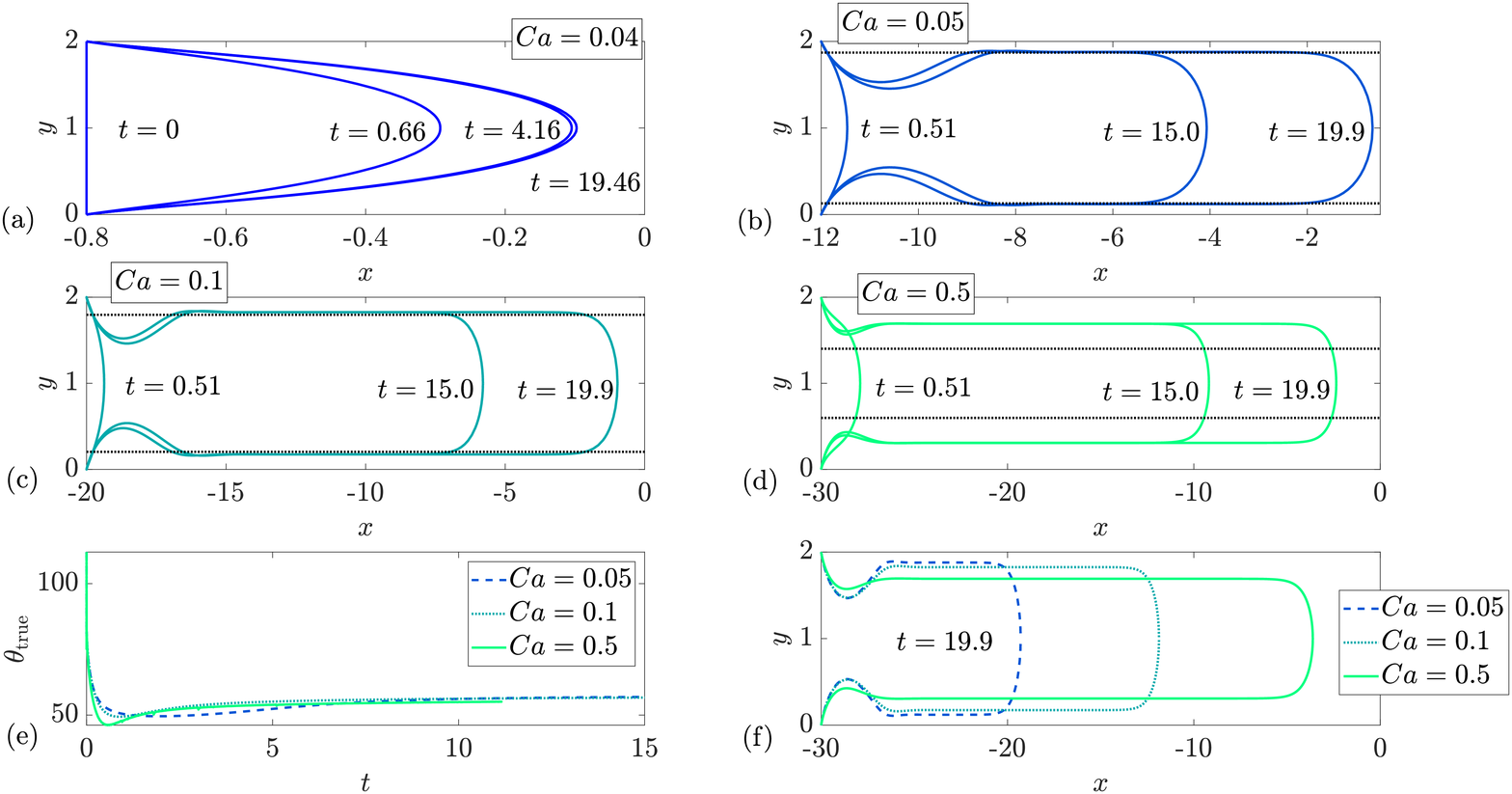}
	\caption{Time-dependent calculation when $\theta_0 = 64.7^\circ$ (or $\lambda = 0.02$). (a) $Ca = 0.04<Ca_{\mathrm{crit}}$. In this case the system settles on the stable steady state and a thin-film is not formed. (b), (c), (d) show the thin-film formation for $Ca = 0.05,0.1,0.5>Ca_{\mathrm{crit}}$ respectively. The dotted lines indicate the Landau-Levich-Derjaguin film height, given in \eqref{LLD}. (e) shows the evolution of $\theta_{\mathrm{cl}}$ as a function of $t$. (f) compares the thin-film profiles for different $Ca$ when $t=19.9$. Note the scale of the horizontal axes on (a)-(d) are different.}
  \label{fig:thin_film_width_1_0}
\end{figure}

\begin{figure}
\centering
	\includegraphics[scale=0.5,trim=0 0 0 0,clip]{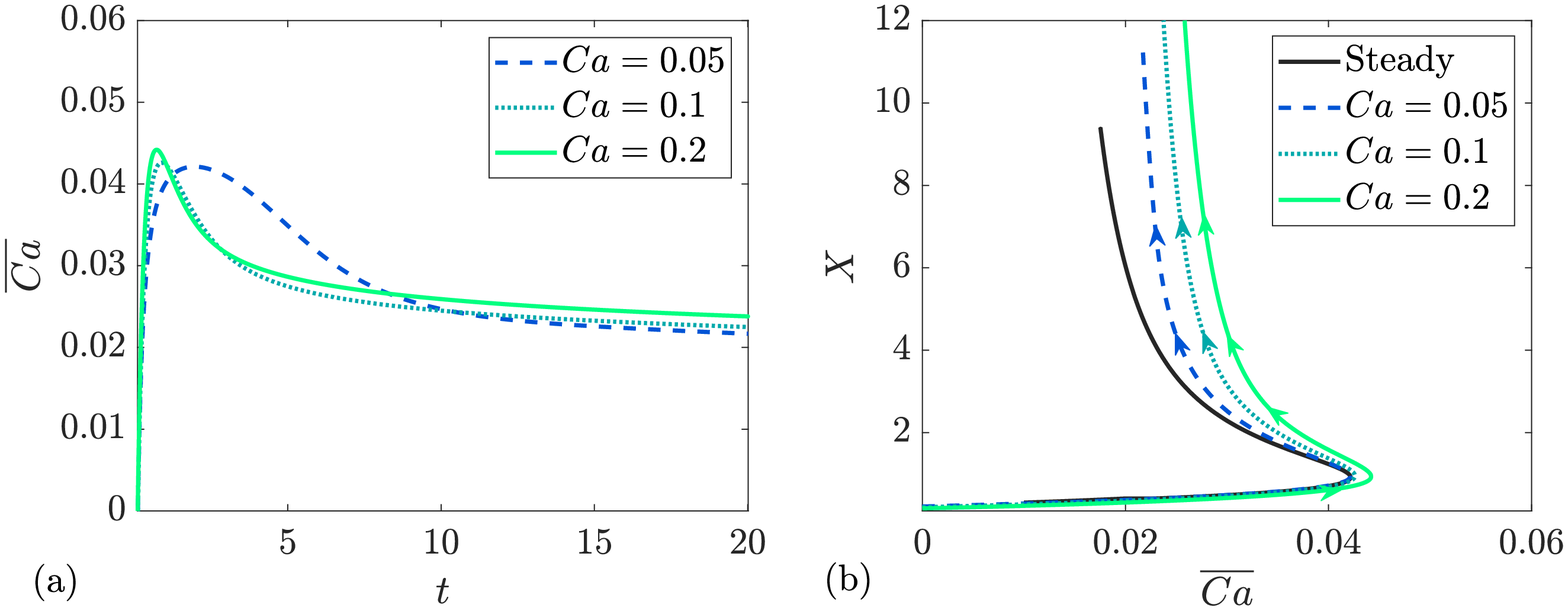}
	\caption{(a) Time-signal of $\overline{Ca}$ defined in \eqref{rel_ca} when $\theta_0 = 64.7^\circ$ (or $\lambda = 0.02$, for $Ca = 0.05,0.1,0.5>Ca_{\mathrm{crit}}$, respectively. (b) Comparison of time trajectories with the steady solution curve in the $(\overline{Ca},X)$ plane.}
  \label{fig:relative_ca}
\end{figure}

We now discuss time-dependent calculations and the formation/deposition of a thin liquid film. In {most of the simulations that follow,} we start a {pressure-driven} system from rest with an initially flat interface and a constant value of $Ca$. As shown in \cite{keeler2021stability}, if we choose $Ca<Ca_{\mathrm{crit}}$ then the system will relax to the stable steady solution branch, as seen in the MD simulations when $F^*_0<F^*_{\mathrm{crit}}$. However if we choose $Ca>Ca_{\mathrm{crit}}$ a thin-film will develop, as also observed in the simulations. In this section we implement \eqref{variable_theta} and perform time-dependent calculations to understand the effect of the various parameters on the formation of this thin-film.

Figure~\ref{fig:thin_film_quiver} is a visualisation of the velocity field using quivers to represent the strength and direction of the flow once a thin-film has developed. There are three distinct regions; a `rim' region close to the contact line, a flat, thin-film region of height $h_{\mathrm{film}}$ and a static region corresponding to the static meniscus shape. We remark that close to the contact line the flow is approximately parallel, and so a lubrication model would be an appropriate model reduction here. Far away from the contact line, the flow is certainly not parallel and, therefore, to resolve the half liquid-plug a full continuum model is required. 

Figure~\ref{fig:thin_film_width_1_0} shows snapshots of the evolution of the interface at different times, $t$, for various values of $Ca$ (panels (a) - (d)). Panel (e) shows the time-signal of $\theta_{\mathrm{cl}}$ and panel (f) compares the final time-snapshot for the different values of $Ca$ chosen. In the super-critical case (i.e. $Ca>Ca_{\mathrm{crit}}$) the height of the thin-film is approximately constant before an almost circular cap region closes the interface. For macroscopic geometries, it is well known that the film thickness, $h_{\mathrm{film}}$ scales according to the Landau-Levich-Derjaguin (LLD) law \citep{landau1988dragging,deryaguin1943thickness}:
\bea
h_{\mathrm{film}} \sim 0.95 Ca^{2/3}, Ca\ll 1.
\label{LLD}
\eea
This value of the film height is shown as dotted lines in figure~\ref{fig:thin_film_width_1_0} (b)-(d) and the actual thin-films closely match this value with increasing accuracy as $Ca$ becomes smaller (as expected).

\begin{figure}
  \centering
  \includegraphics[scale=0.35,trim=0 0 0 0,clip]{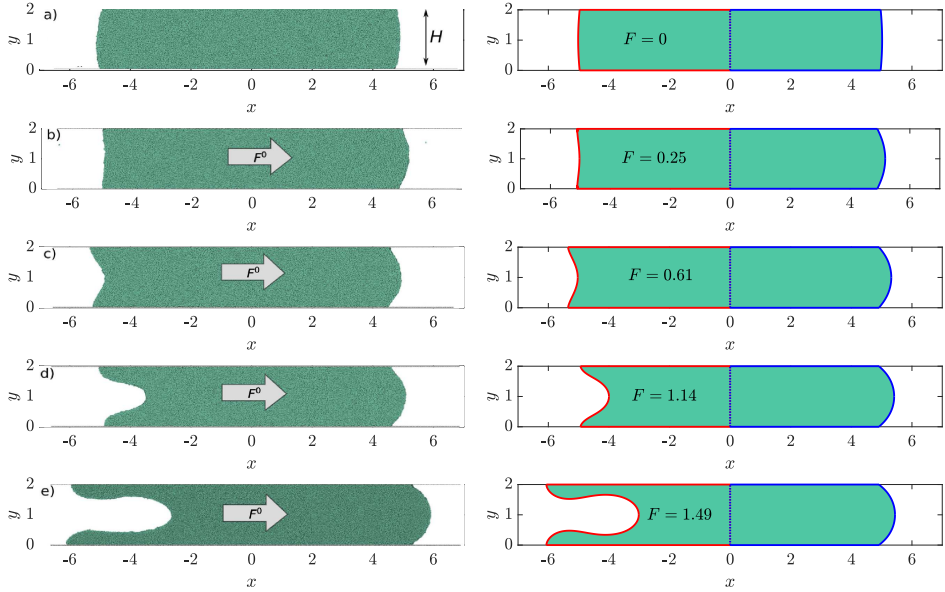}
	\caption{Qualitative comparison with figure~\ref{fig:blake_figure} (Figure reprinted from \cite{toledano2021closer}, with permission from Elsevier.) when $\theta_0 = 102.4^\circ$. The left column are the images taken from figure~\ref{fig:blake_figure} and the right column are calculations using the {force-driven problem with the value of $F$ stated}.}
  \label{fig:blake_comparison}
\end{figure}

The contact angle at small times rapidly decreases and achieves a minimum value, before gradually increasing to a limiting value, at $\theta_{\mathrm{cl}} \approx 59^\circ$, as shown in panel (e); the same time-dependent behaviour was observed in the MD simulations.  A key observation is that in these time-dependent calculations the limiting relative capillary number $\overline{Ca}$ is independent of $Ca$, as seen in figure~\ref{fig:relative_ca}(a) which is consistent with experimental and theoretical studies, for example \cite{snoeijer2006avoid}. This indicates that the flow in the film region becomes increasingly independent of the liquid-plug. In \cite{keeler2021stability} it was shown that, at the RCL especially, the time-dependent trajectories of the system are similar to the steady bifurcation diagram when both are plotted in the $(\overline{Ca},X)$ plane. The same phenomenon occurs here; see panel (b) in figure~\ref{fig:relative_ca} where it is shown the trajectories closely match the steady bifurcation curve. This is consistent with a prediction of \cite{snoeijer2012theory} that, for plate-withdrawal from a bath flattened in the far-field by gravity, the dynamics closely follow the unstable branch of solutions in a quasi-steady manner. In their case, where gravity plays an important role, the bifurcation curve oscillates around a fixed value of $Ca$, but in the pressure-driven problem this does not occur and we observe monotonic convergence towards a particular $Ca$. {In the body-force problem we see the exact same phenomena, for both the CA and VA model, (results not shown), but leave a thorough investigation of this as a future research avenue.} Finally, we can make a qualitative comparison of the numerical results to the MD results in figure~\ref{fig:blake_figure}. Figure~\ref{fig:blake_comparison} shows the equivalent half liquid plug profiles for the force-driven problem (as in the MD) obtained by computing the receding and advancing interfaces separately and combining them. {As can be seen from the profiles, the qualitative comparison is strong.}

\section{Larger Scale Systems\label{sec:large_scale}}

\begin{figure}
  \centering
  \includegraphics[scale=0.35,trim=0 0 0 0,clip]{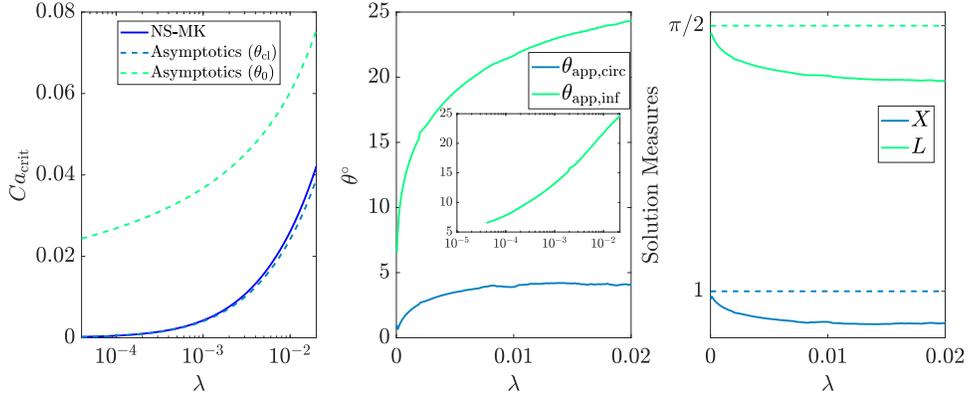}
  \caption{(a) The evolution of $Ca_{\mathrm{crit}}$ as $\lambda$ is varied for $\theta_0 = 64.7^{\circ}$. The numerics are indicated by the solid curve and the asymptotics given in \eqref{ca_crit} are shown with a dashed curve when $\theta_{\mathrm{cl}}$ is used in \eqref{ca_crit} and a dotted curve when $\theta_{0}$ is used in \eqref{ca_crit}. (b) The variation of $\theta_{\mathrm{app,circ}}$ and $\theta_{\mathrm{app,inf}}$ at the critical point, shown in dashed and solid line respectively. The inset diagram is the same data shown on a log scale. (c) The variation of $X$ and $L$ at the critical point as $\lambda$ is varied.}
  \label{fig:large_scale_loci}
\end{figure}

\begin{figure}
  \centering
  \includegraphics[scale=0.3,trim=0 0 0 0,clip]{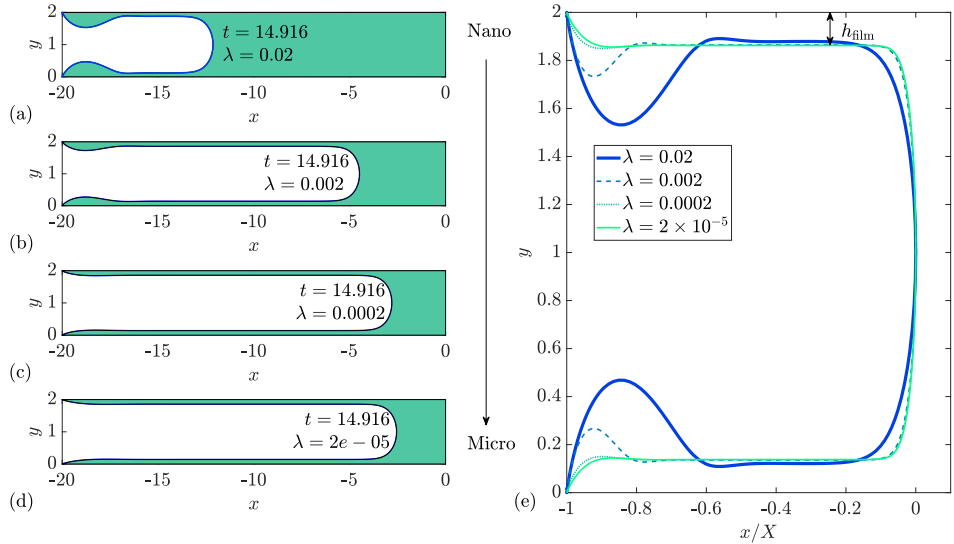}
  \caption{Thin-film formation when the scale of the system is increased. Figures (a) - (d) show the thin film at $t=14.916$ for $\lambda = 0.02,0.002,0.0002,0.00002$ and $Ca = 0.05$, $\theta_0 = 64.7^{\circ}$. Panel (e) shows a comparison of each of the profiles in (a)-(d) scaled by $X$}
  \label{fig:thin_film_comparison}
\end{figure}

Having validated our model in the context of the nano-geometry using MD calculations, we can extend our analysis to investigate thin-film formation in a larger-scale geometry, for which MD simulations are prohibitively computationally expensive. We can achieve this by reducing the value of the dimensionless slip-length $\lambda$ while keeping $\theta_0$, and therefore the physical slip-length constant, which has the effect of increasing $H^*$, the physical channel width. 

We also investigate the formation of thin-films in the limit as $\lambda\to 0$ (we note that $\lambda = 0$ has no solution \citep{huh1971hydrodynamic}). Using the same methods as before, we can track $Ca_{\mathrm{crit}}$ as $\lambda$ is varied. Figure~\ref{fig:large_scale_loci}(a) shows that $Ca_{\mathrm{crit}}\to 0$ as $\lambda\to 0$, indicating that the system becomes unstable for increasingly slower wall speeds as the scale of the system is increased. The dashed line indicates the asymptotic formula given in \eqref{ca_crit}, with the full expression for $\theta_{\mathrm{cl}}$ used, and the dotted line shows \eqref{ca_crit} with $\theta_{\mathrm{cl}}$ replaced with $\theta_0$. {Because $\lambda\ll \delta$, the difference between $\theta_0$ and $\theta_{\mathrm{cl}}$ is large, see equation \eqref{variable_theta}. As a result, in order to capture the numerics, the full expression for $\theta_{\mathrm{cl}}$ has to be included in the asymptotic formula, and then the agreement is excellent. }


Panel (b) shows how $\theta_{\mathrm{app,circ}}$ and $\theta_{\mathrm{app,inf}}$ vary at the critical point. It is clear that for both measures of the apparent angle, as $\lambda$ decreases, $\theta_{\mathrm{app}}\to 0$. This is an important observation and provides a link to the work of \cite{snoeijer2007part1,snoeijer2006avoid,eggers2004forced}, where in the lubrication approximation they apply it is perfectly reasonable to apply the Cox-Voinov formula with $\theta_{\mathrm{app}}=0$ as a means of determining $Ca_{\mathrm{crit}}$. In a nano-geometry however, this approximation is not valid, as we have shown that $\theta_{\mathrm{app,crit}}\neq 0 $. We also find that as $\lambda \to 0$, the interface approaches a circular meniscus with radius $1$ and length $\pi/2$, as the results in panel (c) clearly show.

We now turn our attention to time-dependent results in the limit as $\lambda \to 0$ with $\theta_0$ being kept fixed. Figure~\ref{fig:thin_film_comparison} shows the thin-film at $t=14.916$ for values of $\lambda = 2\times 10^{-2},2\times 10^{-3},2\times 10^{-4},2\times 10^{-5}$ (panels (a) - (d)) with $Ca = 0.05$, $\theta_0 = 64.7^{\circ}$. The largest value of $\lambda = 0.02$ corresponds to the nano-channel considered in \cite{toledano2021closer} while the smallest value of $2\times 10^{-5}$ corresponds to a system $10^3$ times larger, i.e. a micro-channel. Panel (e) shows the comparison of the profiles when the $x$ normalised by $X$. The immediate observation is that the dimensionless film-height, $h_{\mathrm{film}}$ is independent of $\lambda$ once the thin-film has had sufficient time to develop, so that the physical film height will scale linearly with the system size. This is especially evident when comparing the interface profiles for $\lambda = 2\times 10^{-3},2\times 10^{-4},2\times 10^{-5}$. Thus, sufficiently far away from the contact line the structure of the thin-film is independent of the size of the geometry (in physical systems this is measured relative to the physical width of the channel). The `rim' region is however highly dependent on $\lambda$ and the details of the contact line angle (see \cite{flitton2004surface}); the smaller the system (larger $\lambda$) the larger the `rim' near the contact line. Therefore, the physical rim height will increase slower than linearly as system size is increased. 

\section{Conclusion}

We have developed a novel molecularly-augmented continuum model, based on a variable true contact angle, that describes the dynamics of a liquid bridge between two parallel plates, and, more generally, describes the RCL and ACL physics. By solving the resulting set of equations numerically, we are able to interpret the maximum speed of dewetting as a fold bifurcation in the steady bifurcation diagram. We find that the maximum speed of wetting $Ca_{\mathrm{crit}}$, calculated as a function of $\theta_{\mathrm{cl}}$, is qualitatively similar to the MD simulations described in \cite{toledano2021closer} and that the estimate of $L^*/L_m^*$ is in excellent agreement.


As well as showing good agreement with the MD simulation, the advantages of this approach is that by replacing the assumption that $\theta_{\mathrm{cl}}$ is constant with the constraints
\begin{equation}
   \begin{split} 
   \overline{Ca} &=\frac{\lambda}{\delta}\left(\cos(\theta_{\mathrm{cl}}) - \cos(\theta_0)\right)\\ \lambda_{\mathrm{MD}}^* &= a\exp\left[b(1 + \cos(\theta_0)\right]
   \end{split}
    \label{variable_angle2},
\end{equation}
the issue of deciding what $\theta_{\mathrm{cl}}$ should be in any hydrodynamic calculation is removed, as it is naturally determined, through \eqref{variable_angle2}, as part of the solution. Furthermore, whereas in previous approaches the slip-length and $\theta_{\mathrm{cl}}$ had to be specified as control parameters, in this model the only hydrodynamic parameter we have to specify is the slip length. With this parameter being difficult to measure, invariably it has been used as an additional fitting parameter that can cover-up for inaccuraces in the constant angle model. We do however have to estimate $\delta$, the width of the TPZ from MD simulations and this provides an additional parameter that has to be known in advance, although this parameter can be far more accurately specified than slip lengths. The comparison between the MD simulations and the VA model is strong, and although some of the physics present in the MD calculations are absent, for example the disjoining pressure, we conclude that the VA model contains the minimum ingredients required to replicate the physics contained in the MD calculations, at least before the thin-film ruptures (see, for example \cite{kreutzer2018evolution,zhao2018forced}).

Our results also illuminate the values of $\theta_{\mathrm{cl}}$ and $\theta_{\mathrm{app}}$ when a partially wetted substrate is withdawn from a pool of liquid at capillary numbers greater than $Ca_{\mathrm{crit}}$. Experiments have shown that attempts at forced dewetting cause the (three-dimensional) contact line to slant at an angle relative to the direction of withdrawal, such that the capillary number in the direction normal to the contact line remains constant at $Ca_{\mathrm{crit}}$. These observations of avoided critical behaviour led to the postulate of a maximum speed of dewetting \citep{Blake1979}. Presumably, $\theta_{\mathrm{cl}}$ and $\theta_{\mathrm{app}}$ along the slanted contact line are the smallest possible consistent with a stable flow without film deposition, i.e. those associated with the turning point in the steady phase diagram. For $\theta_{\mathrm{cl}}$ this is $\theta_{\mathrm{cl,crit}}$. For $\theta_{\mathrm{app}}$ it depends on how the angle is measured. 

We are easily able to extend the VA model to larger systems, which are prohibitively computationally expensive for MD calculations, and by examining the thin-film formation in these systems when $Ca>Ca_{\mathrm{crit}}$, we are able to demonstrate that the relative height of the thin-film is independent of size of the system and weakly dependent on $Ca$. Differences in the interface profile occur close to the contact line, as indicated by the size of the `rim' that develops, but sufficiently far away from the contact line the relative heights of the thin-film are nearly identical.

Another advantage of the framework is practical, in that the computational time for these calculations is $O(\mathrm{minutes})$ using the open-source \texttt{oomph-lib} framework with state of the art linear algebra solvers, rather than $O(\mathrm{days})$ for the MD simulations. As we are able to obtain the velocity and pressure fields in addition, this unified model has excellent potential for researchers wishing to combine the best aspects of the hydrodynamic and molecular theories in their work. We also remark that viscous and inertial effects can be incorporated in this model by, for example, treating the gas-phase using a lubrication approximation; see \cite{keeler2021stability}. We also remark that the QP model and associated asymptotic results, while not resolving the flow-field, are useful for validation, as demonstrated here.

Nevertheless, there remains a need for more physical experiments with emphasis on the RCL up to the point of film deposition, since, as we have seen, this encodes much valuable information concerning the contact angle on the microscopic scale.  While there is a very large body of literature on film deposition, such as that which occurs when a solid surface is withdrawn from a pool of liquid, and much published data on advancing contact angles, comprehensive measurements of dynamic receding angles on partially-wetted surfaces are, unfortunately, rare. A resurgence of interest is overdue.

\appendix

\section{MKT Theory \label{app:MD_theory}}

According to the molecular-kinetic theory of dynamic wetting (MKT), the contact line advances or recedes across the energy landscape of the solid surface as a consequence of random, thermally-activated molecular events having characteristic frequency $\kappa_0^*$ and length $\lambda_0^*$ (not to be confused with the dimensionless slip-length $\lambda$) \citep{Blake1969,Blake1993}. Such events occur across the whole solid-liquid interface, but only those that that take place within the three-phase zone (TPZ) determine dynamic wetting. The TPZ, of width $\delta^*$, is the region where the liquid-vapour and solid-liquid interfaces meet, i.e. the contact line viewed at the molecular scale.  At equilibrium, the molecular events simply cause the contact line to fluctuate about its mean position \citep{tolendano2019,toledano2020hidden,tolendano2020b}.  However, for net displacement of the contact line at velocity $U_{\mathrm{cl}}^*$, work must be done to favour events in the desired direction.  This work is provided by the out-of-balance surface tension force that arises when the equilibrium is disturbed: $f^* = \gamma_L^*(\cos(\theta_0) - \cos(\theta_{\mathrm{cl}}))$, where $\gamma_L^*$ is the surface tension of the liquid. According to the model, as the TPZ moves across the solid surface, this work is expended at $n^*$ interaction sites per unit area swept.  Application of the Frenkel-Eyring theory of stress-modified activated rate processes \citep{Frenkel1946,Glasstone1941} then leads to the principal equation linking $U_{\mathrm{cl}}^*$ and $\theta_{\mathrm{cl}}$:
  \bea
  U_{\mathrm{cl}}^* = 2\kappa_0^*\lambda_0^*\sinh\left[\gamma_L^*\left(\cos(\theta_0) - \cos(\theta_{\mathrm{cl}})\right)/2n^*\,k_B\,T\right],
  \eea
  where $k_B$ and $T$ are, respectively, the Boltzmann constant and the absolute temperature.

  Since its inception, this equation has proved very effective in correlating experimental and MD data for a wide range of systems.  For examples see \cite{Blake1993,Schneemilch1998,Blake2006,Duvivier2013}.  In the interpretation of experimental data, the interactions sites are usually assumed to be uniformly distributed, so that $\lambda_0^*\approx 1/\sqrt{n^*}$; thus, reducing the unknowns to just two: $\kappa_0^*$ and $\lambda_0^*$.

  For small arguments of $\sinh$, typically when $\theta_{\mathrm{cl}}$ is not too far from $\theta_0$ (true for $\theta_{\mathrm{cl}}$ at the RCL in the MD data investigated here) or $\gamma_L^*$ is small, this reduces to a linear relationship:
  \bea
  U_{\mathrm{cl}}^* = \left(\frac{\kappa_0^*\lambda_0^*}{n^*\,k_B\,T}\right)\gamma_L^*\left(\cos(\theta_0) - \cos(\theta_{\mathrm{cl}})\right),
  \eea
  which may be written as
  \bea
  Ca_{\mathrm{cl}} = \frac{\mu_{L}^*}{\zeta^*}\left(\cos(\theta_0) - \cos(\theta_{\mathrm{cl}})\right),
  \label{appendix1_equation}
  \eea
  where $\zeta^*$ is the coefficient of contact-line friction (per unit length of the contact line):
  \bea
  \zeta^* = \frac{n^*\,k_B\,T}{\kappa_0^*\,\lambda_0^*}.
  \eea
  This single coefficient quantifies the localised resistance to the displacement of the contact line.

  In previous MD studies \citep{Blake2015,Bertrand2009} it has been shown that both contact-line friction and slip between a liquid and a solid depend on the same thermally-activated molecular events.  Whereas, at the contact line, the principle driving force comes from the out-of-balance surface tension acting across the TPZ, for the latter it is provided by the viscous shear stress acting across the whole solid-liquid interface: $\mu_L^*\left(\partial u^*/\partial z^*\right) = \beta^*\lambda^*$, where $\beta^*$ is the slip coefficient and $\lambda^*$ the Navier slip length (i.e. the distance into the solid at which the extrapolated fluid velocity vanishes).  Because of the common mechanism, it follows that the two coefficients are directly related; specifically, 
  \bea
  \beta^* = \zeta^*/\delta^*;
  \eea
  hence,
  \bea
  \lambda^* = \delta^*\mu_L^*/\zeta^*.
  \label{appendix2_equation}
  \eea
This relationship has been validated by molecular-dynamics simulations, in which both the contact-line friction and the slip length have been measured for the same system over a range of equilibrium contact angles \citep{Blake2015,toledano2020hidden}.  Good agreement has been shown for both Lennard-Jones liquids and atomistically simulated water on molecularly smooth carbon-like surfaces.  That said, a precise correlation hinges on the value of $\delta^*$. For the Lennard-Jones liquids, the value selected was assessed from the velocity profiles across the TPZ.  For the simulated water system, the distance over which the density of the of the liquid in contact with the solid fell to zero was used.  See figure 10 in~\cite{Blake2015} to compare the two approaches.  Arguments may be made for both.  For the Lennard-Jones system, the difference in the result was in the region of 30\%.  Slip lengths were smaller if the density profile was used.  In addition, the value of $\delta^*$ appeared to depend weakly on both contact-line velocity and the equilibrium contact angle. Based on the existing data, while \eqref{appendix2_equation} appears to be physically justified, a precise understanding of the subtle influences in play requires more work.  The value of $\delta^*$ found for the coarse-grained water simulations \citep{toledano2021closer} was 0.93 ± 0.14 nm based on the density argument.  We use this value in the present paper.

If, \eqref{appendix2_equation} is accepted, at least in principle, it allows us to rewrite \eqref{appendix1_equation} in dimensionless variables, as 
\bea
Ca_{\mathrm{cl}} = \frac{\lambda}{\delta}\left(\cos(\theta_0) - \cos(\theta_{\mathrm{cl}})\right).
\eea
In the stationary frame of the liquid-plug between two solid walls moving at velocity $U^*_{\mathrm{wall}}$, this becomes
\bea
Ca\left(U_{\mathrm{wall}} - \pdiff{x_{\mathrm{cl}}}{t}\right) = \frac{\lambda}{\delta}\left(\cos(\theta_0) - \cos(\theta_{\mathrm{cl}})\right),
\eea
which is \eqref{variable_theta} in the main body of the paper when we set the nondimensional wall speed to be $U_{\mathrm{wall}} = -1$. Furthermore, and perhaps more significantly, we know that contact-line friction depends strongly on the equilibrium contact angle.  This means that the same is true for the slip length.  As has been shown \cite{Blake1993,Blake2002,Bertrand2009,Duvivier2013} the frequency $\kappa_0^*$  is related to the equilibrium contact angle by 
\bea
\kappa_0^*\sim\left(k_B\,T/\mu_L^*v_L^*\right)\exp\left[-\gamma_L^*\left(1 + \cos(\theta_0)\right)/n^*\,k_B\,T\right],
\eea
where $v_L^*$ is the molecular flow volume in the Frenkel-Eyring theory. This leads to
\bea
\zeta^*\sim\left(n^*\mu_L^*v_L^*/\lambda_0^*\right)\exp\left[-\gamma_L^*\left(1 + \cos(\theta_0)\right)/n^*\,k_B\,T\right]
\eea
and, hence, to
\bea
\lambda^*\sim\delta^*\left(\lambda_0^*/n^*v_L^*\right)\exp\left[-\gamma_L^*\left(1 + \cos(\theta_0)\right)/n^*\,k_B\,T\right].
\eea
This suggests the general (dimensionless) form
\bea
\lambda = a\exp\left[b(1 + \cos(\theta_0)\right].
\eea
In the present paper we have used this expression,  \eqref{lambda_theta_0}, to fit the slip length calculated from the MD data in table 1 of \cite{toledano2021closer}.

\vskip2pc

\bibliographystyle{jfm} 

\bibliography{dynamic_wetting1}

\end{document}